%% file: main.tex
\documentclass[aip,reprint]{revtex4-2}

\usepackage[utf8]{inputenc} 
\usepackage[T1]{fontenc}    
\usepackage{hyperref}       
\usepackage{url}            
\usepackage{booktabs}       
\usepackage{amsfonts}       
\usepackage{nicefrac}       
\usepackage{microtype}      
\usepackage{graphicx}
\usepackage{lipsum}
\usepackage{comment}
\usepackage[caption=false]{subfig}
\usepackage{textcomp}
\usepackage{soul}           
\usepackage{amsmath}
\usepackage{listings}
\usepackage{chngcntr}
\usepackage{enumitem}
\usepackage[table,xcdraw]{xcolor}
\usepackage{booktabs}
\usepackage{multirow}
\usepackage{hhline}

\newcommand{\yuanpeng}[1]{\textcolor{black}{#1}}

\definecolor{codegreen}{rgb}{0,0.6,0}
\definecolor{codegray}{rgb}{0.5,0.5,0.5}
\definecolor{codepurple}{rgb}{0.58,0,0.82}
\definecolor{backcolour}{rgb}{0.95,0.95,0.92}

\lstdefinestyle{mystyle}{
    backgroundcolor=\color{backcolour},   
    commentstyle=\color{codegreen},
    keywordstyle=\color{magenta},
    numberstyle=\tiny\color{codegray},
    stringstyle=\color{codepurple},
    basicstyle=\ttfamily\footnotesize,
    breakatwhitespace=false,         
    breaklines=true,                 
    captionpos=b,                    
    keepspaces=true,                 
    numbers=left,                    
    numbersep=5pt,                  
    showspaces=false,                
    showstringspaces=false,
    showtabs=false,                  
    tabsize=2
}
\let\oldtextbf=\textbf
\renewcommand*{\textbf}[1]{\ifmmode\mathbf{#1}\else\oldtextbf{#1}\fi}
\renewcommand*{\phi}[0]{\varphi}

\draft 

\begin{document}

\title[]{The role of fluctuations in the nucleation process} 



\author{Yuanpeng Deng}
\affiliation{ 
Key Lab of Smart Prevention and Mitigation of Civil Engineering Disasters of the Ministry of Industry and Information Technology and Key Lab of Structures Dynamic Behavior and Control of the Ministry of Education, Harbin Institute of Technology, 150090 Harbin, China%
}%
\affiliation{Atomistic Simulations, Italian Institute of Technology, 16156 Genova, Italy}

\author{Peilin Kang}
\affiliation{Atomistic Simulations, Italian Institute of Technology, 16156 Genova, Italy}

\author{Xiang Xu}%
\altaffiliation{xuxiang@hit.edu.cn}
\affiliation{ 
Key Lab of Smart Prevention and Mitigation of Civil Engineering Disasters of the Ministry of Industry and Information Technology and Key Lab of Structures Dynamic Behavior and Control of the Ministry of Education, Harbin Institute of Technology, 150090 Harbin, China%
}%

\author{Hui Li}%
\altaffiliation{lihui@hit.edu.cn}
\affiliation{ 
Key Lab of Smart Prevention and Mitigation of Civil Engineering Disasters of the Ministry of Industry and Information Technology and Key Lab of Structures Dynamic Behavior and Control of the Ministry of Education, Harbin Institute of Technology, 150090 Harbin, China%
}%

\author{Michele Parrinello}
\email[]{michele.parrinello@iit.it}
\affiliation{Atomistic Simulations, Italian Institute of Technology, 16156 Genova, Italy}


\date{\today}

\begin{abstract}
\input{manuscript/abstract}

\end{abstract}


\maketitle 

\input{manuscript/paper}



%
%

%

\begin{acknowledgments}
The authors want to acknowledge Jintu Zhang and Enrico Trizio for the many helpful conversations.
This study was financially supported by the National Natural Science Foundation of China (Grant Nos. 51921006, 52192661, and 52322803).
\end{acknowledgments}

\section*{Code and Data Availability} \label{sec:code_avail}
    The NN-based committor models are based on the Python machine learning library PyTorch.~\cite{paszke_pytorch_2019}
    The specific code for the definition and the training of the model is developed in the framework of the opensource \texttt{mlcolvar}~\cite{bonati_unified_2023} library. 
    The committor-based enhanced sampling simulations have been performed using LAMMPS~\cite{thompson_lammps_2022} patched with PLUMED~\cite{tribello_plumed_2014} 2.9.
    All the reported molecular snapshots have been produced using the OVITO.~\cite{stukowski_visualization_2010}

\section*{Bibliography}
\bibliography{references/1Nucleation}


\setcounter{section}{0}
\renewcommand{\thesection}{S\arabic{section}}
\setcounter{equation}{0}
\renewcommand{\theequation}{S\arabic{equation}}
\setcounter{figure}{0}
\renewcommand{\thefigure}{S\arabic{figure}}
\setcounter{table}{0}
\renewcommand{\thetable}{S\arabic{table}}
    
\clearpage
\onecolumngrid

{\Large\normalfont\sffamily\bfseries{{Supporting Information}}}

\input{manuscript/supporting}


\end{document}

%% file: manuscript/abstract.tex
The emergence upon cooling of an ordered solid phase from a liquid is a remarkable example of self-assembly, which has also major practical relevance.
Here, we use a recently developed committor-based enhanced sampling method [Kang \textit{et al.}, Nat. Comput. Sci. \textbf{4}, 451\text{-}460 (2024); Trizio \textit{et al.}, Nat. Comput. Sci. 1\text{-}10 (2025)] to explore the  crystallization transition in a Lennard-Jones fluid, using Kolmogorov's variational principle. 
In particular, we take advantage of the properties of our sampling method to harness a large number of configurations from the transition state ensemble. 
From this wealth of data, we achieve precise localization of the transition state region, revealing a nucleation pathway that deviates from idealized spherical growth assumptions. 
Furthermore, we take advantage of the probabilistic nature of the committor to detect and analyze the fluctuations that lead to nucleation.
Our study nuances classical nucleation theory by showing that the growing nucleus has a complex structure, consisting of a solid core surrounded by an interface that is more disordered than bulk liquid.
We also compute from the Kolmogorov's principle a nucleation rate that is consistent with the experimental results at variance with previous computational estimates. 

%% file: manuscript/paper.tex

Nucleation is a fundamental physical process which underpins a wide range of phenomena, from the formation of crystals in industrial applications to natural events such as ice formation.~\cite{kelton_crystal_1991,oxtoby_homogeneous_1992}
Our understanding of the phenomenon is still largely based on classical nucleation theory (CNT), which describes the process as due to fluctuations in the undercooled liquid phase in which nuclei of the ordered phase are fleetingly formed.
For such fluctuations to eventually lead to crystallization, the size of the nucleus must be such that the free energy gain of the atoms in the crystallite balances the free energy cost of forming an interface between crystallite and liquid.
The set of configurations for which this balance occurs identifies the transition state ensemble (TSE).

While in general terms this picture is accepted, the details of how this transition takes place are not fully understood and the fluctuations that drive this phenomenon have not yet been described. 
Unfortunately, the process of nucleation is difficult to investigate experimentally since it involves studying nanometer scale phenomena that take place in a very short time at random intervals.
Molecular dynamics (MD) simulations are in principle well positioned to probe the early stages of nucleation.~\cite{mandell_crystal_1977,stillinger_study_1978,hsu_crystal_1979}
However, nucleation is a rare event that takes place on time scales that cannot be reached by brute-force MD simulations.~\cite{giberti_metadynamics_2015,sosso_crystal_2016,finney_molecular_2024}
This obstacle has not stopped researchers, and a variety of enhanced sampling methods have helped alleviating this problem.~\cite{torrie_nonphysical_1977,laio_escaping_2002,invernizzi_rethinking_2020,invernizzi_exploration_2022,dellago_transition_1998,van_erp_novel_2003,allen_forward_2009,espinosa_seeding_2016}
Thanks to these simulations, much is known about this transition especially when it comes to its prototypical model system, i.e., the nucleation of a Lennard-Jones (LJ) liquid.~\cite{rein_ten_wolde_numerical_1996,moroni_interplay_2005, trudu_freezing_2006, jungblut_crystallization_2011}
Still, as we shall show below, these analyses have been performed on a relatively small set of crystallization trajectories. 

Some researchers, have rightly focused their attention on the committor $q(\mathbf{x})$,~\cite{kolmogoroff_uber_1931} which is the quantity that most rigorously describes rare events. 
We recall that given two metastable basins $A$ and $B$, the function $q(\mathbf{x})$ is the probability that a trajectory passing through configuration $\mathbf{x}$ ends in $B$ without having first visited $A$.~\cite{e_transition-path_2010}
Furthermore, computing $q(\mathbf{x})$ allows defining the TSE and it is believed to be the best possible collective variable \yuanpeng{(CV)}.~\cite{berezhkovskii_one-dimensional_2005, ma_automatic_2005, li_recent_2014, he_committor-consistent_2022}
Unfortunately, the standard way of computing $q(\mathbf{x})$ involves launching a large number of trajectories, waiting for them to end either in $A$ or $B$ and evaluating from the statistics thus gathered the probability $q(\mathbf{x})$.~\cite{moroni_interplay_2005, trudu_freezing_2006, jungblut_crystallization_2011}
In addition to being rather lengthy, this procedure may depend on the limits that are set to the time that one waits before declaring the trajectory committed to either $A$ or $B$.~\cite{lazzeri_molecular_2023}

This complex procedure is in principle not necessary.
In fact, it has been shown that \yuanpeng{$q(\mathbf{x})$} can be computed using a variational principle due to Kolmogorov.~\cite{vanden-eijnden_transition_2005,maragliano_temperature_2006,zinovjev_transition_2015,khoo2019solving,li2019computing} 
Very recently, we have set up an self-consistent procedure that is extremely efficient in computing not only the committor~\cite{kang_computing_2024} but also the free energy,~\cite{trizio_everything_2025} and like Ref.~\cite{vanden-eijnden_transition_2005,maragliano_temperature_2006,zinovjev_transition_2015} \yuanpeng{only requires the information of initial and final states.}

The Kolmogorov's principle states that $q(\textbf{x})$ can be computed by minimizing the functional $\mathcal{K}[q(\textbf{x})]$,~\cite{kolmogoroff_uber_1931}

    \begin{equation}
        \mathcal{K}[q(\textbf{x})] =\Big \langle \big|\nabla_\mathbf{u} q(\textbf{x})\big|^2 \Big\rangle_{U(\textbf{x})} 
        \label{eq:variational_functional}
    \end{equation}

\noindent where the average is over the Boltzmann distribution associated with the interatomic potential $U(\textbf{x})$ at the inverse temperature $\beta$, $\nabla_\textbf{u}$ denotes the gradient with respect to the mass-weighted coordinates, and the boundary conditions $q({\textbf{x}}_A)=0$ and $q({\textbf{x}}_B)=1$ for configurations $\textbf{x}_A$ and $\textbf{x}_B$ in the two basins are imposed.

    \begin{figure}[]
        \centering
            \includegraphics[width=0.95\linewidth]{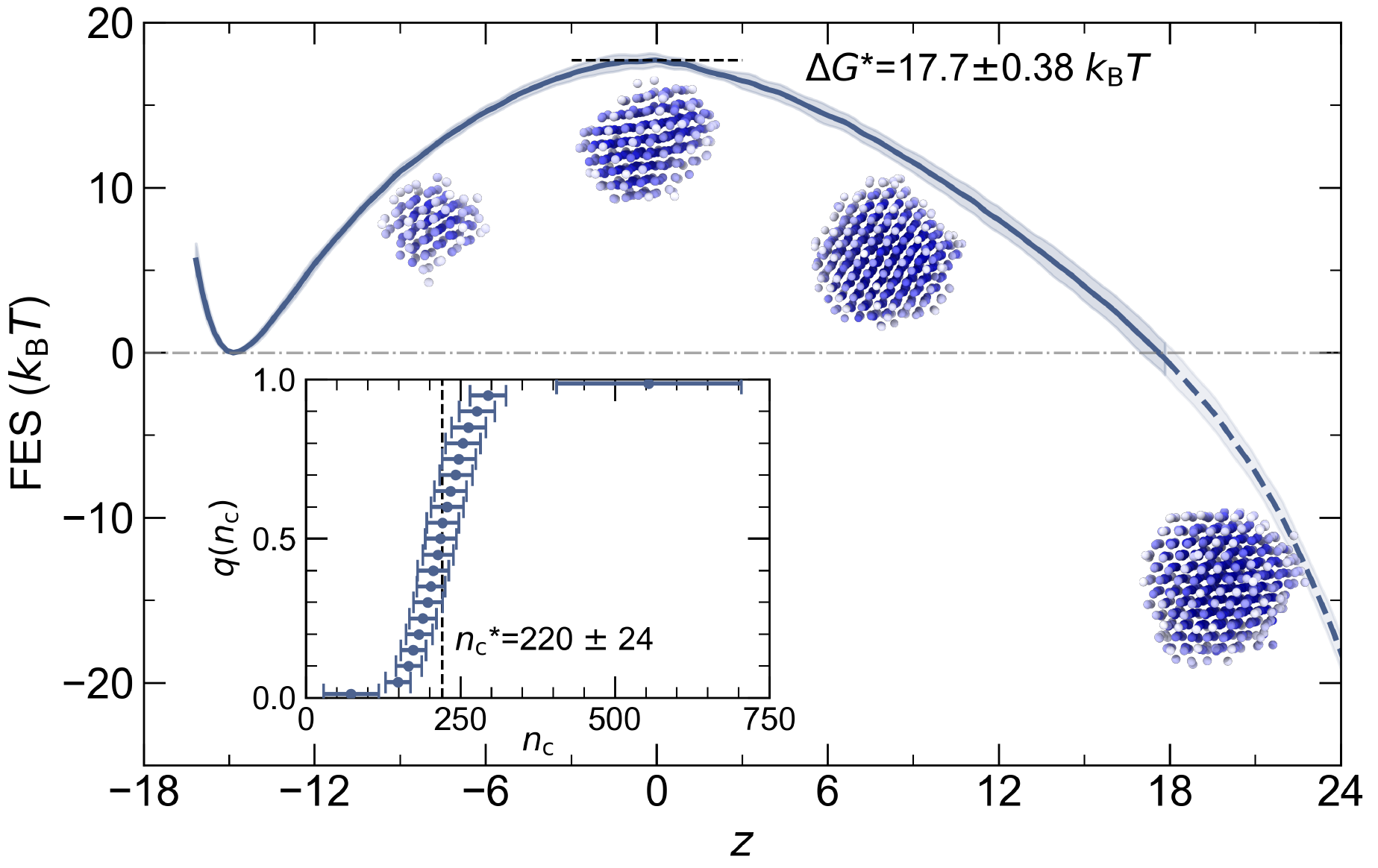}
        \caption{Free energy as function of $z$ for a 2916-atom Lennard-Jones (LJ) system at 0.725 $T_m$.
        The standard deviation over four different OPES+$V_\mathcal{K}$ simulations is also shown as a shaded region.
        We signal with a dashed line the region for $z > 18$ where the results might have been influenced by the use of the radius restraint.
        Snapshots of representative during the nucleation process for different values of $z$, the largest one is taken from the $B$ state.
        The inset shows the variation of the committor function $q(n_c)$ projected on the number of atoms in the largest crystalline cluster, $n_\text{c}$, as defined in Ref.~\cite{rein_ten_wolde_numerical_1996}.} 
        \label{fig:1}
    \end{figure}

\yuanpeng{In practice, estimating $q(\mathbf{x})$ remains challenging even with variational methods.
In a rare event scenario $q(\mathbf{x})$ transitions sharply from $q(\mathbf{x}) \approx 0$ for $\mathbf x \in A$ to $q(\mathbf{x}) \approx 1$ for $\mathbf x \in B$, thus $\big|\nabla_\mathbf{u} q(\textbf{x})\big|^2 $ is strongly peaked at the transition state (TS) region, the very region that is difficult to sample.
To accurately estimate and minimize $\mathcal{K}[q(\mathbf{x})]$, we introduced in Ref.~\cite{kang_computing_2024} a bias potential dependent on $q(\textbf{x})$ that effectively focuses sampling on the TS region:}

    \begin{equation}
        V_\mathcal {K}(\textbf{x})=-\frac{1}{\beta}\log (|\nabla q(\textbf{x})|^2)
    \label{eq:bias_potential_Vk}
    \end{equation}

However, since this bias depends on the solution of the variational principle, this appears to be just an academic observation.
Luckily, this chicken and egg problem can be solved using a \yuanpeng{self-consistent} iterative procedure.
\yuanpeng{Direct use of $q(\mathbf{x})$ as a CV is problematic due to its near-constant values in the metastable basins and sharp variations in the TS region, leading to numerical difficulties.\cite{khoo2019solving,trizio_everything_2025}
In Ref.~\cite{trizio_everything_2025} we use the logarithmic transformation of the committor $z(\textbf{x})=\log{\big(\frac{q(\textbf{x})}{1-q(\textbf{x}\big )}\big)}$ as the CV to drive exploration in the Kolmogorov ensemble $U(\mathbf{x}) + V_\mathcal{K}(\mathbf{x})$.
Combined with On-the-fly Probability Enhanced Sampling (OPES),\cite{invernizzi_rethinking_2020,invernizzi_exploration_2022} this approach allows a thorough and balanced sampling of both metastable and transition states.}

In Ref.~\cite{kang_computing_2024,trizio_everything_2025}, we have also extended the usual notion of TSE which is defined as the probability distribution:

    \begin{equation}
       p_\mathcal{K}(\mathbf{x})=\frac{|\nabla_\mathbf{u} q(\textbf{x})|^2 e^{-\beta U(\mathbf {x})}}{Z_\mathcal{K}}   
    \label{eq:Kolmogorov_probability_pk}
    \end{equation}

\noindent where $Z_\mathcal{K}={\int e^{-\beta U_\mathcal{K}(\mathbf{x})} \, \mathrm{d}\mathbf{x}}$ is a normalizing constant.
The rational for this choice is that $p_\mathcal{K}(\mathbf x)$ is proportional to the contribution that the trajectories passing via $\mathbf x$ gives to the transition rate.
This leads to a more nuanced description of the TSE than the standard $q(\mathbf{x}) \simeq 0.5$ one. 

In this paper we demonstrate that our new method~\cite{kang_computing_2024,trizio_everything_2025} is able to tackle efficiently a real-life nucleation problem.
In particular, we study here a Lennard-Jones system whose $\epsilon$ and $\sigma$ values have been fixed to values appropriate to argon.
\yuanpeng{The unprecedented amount of data collected on the TS region and the probabilistic nature of the transition have allowed us to put on firmer statistical ground the fact that the critical nucleus is non-spherical.
In addition, our analysis regarding the liquid fluctuations along the nucleation process shows that the liquid surrounding the growing nucleus is less ordered than the bulk liquid.}
Finally we show how from the Kolmogorov's approach one can determine the nucleation rate of the system.

    \begin{figure}[b]
        \centering
           \includegraphics[width=0.95\linewidth]{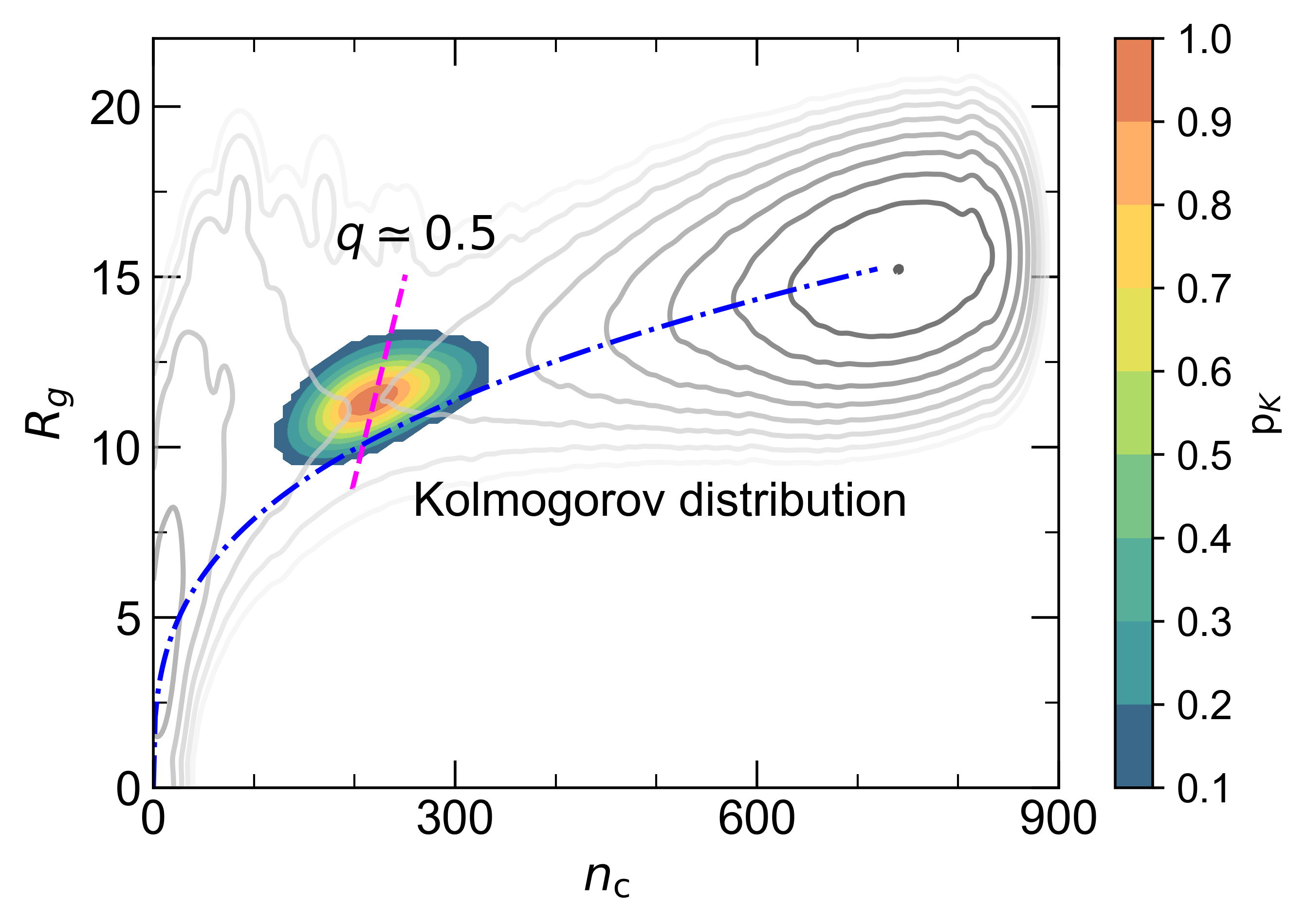}
        \caption{Distribution density of TSE (see Eq.~\ref{eq:Kolmogorov_probability_pk}) projected in the $n_\text{c}$, $R_g$ plane.
        The gray lines are free energy levels, the magenta dashed line shows the isocommittor line $q({n_c,R_g}) = 0.5$.
        The blue dashed-dotted line shows the size-scaling $R_g \propto n_\text{c}^{1/3}$ where the proportionality constant is fitted to state $B$.}
        \label{fig:2}
    \end{figure}

    \begin{figure*}[htbp!]
        \centering
           \includegraphics[width=1\linewidth]{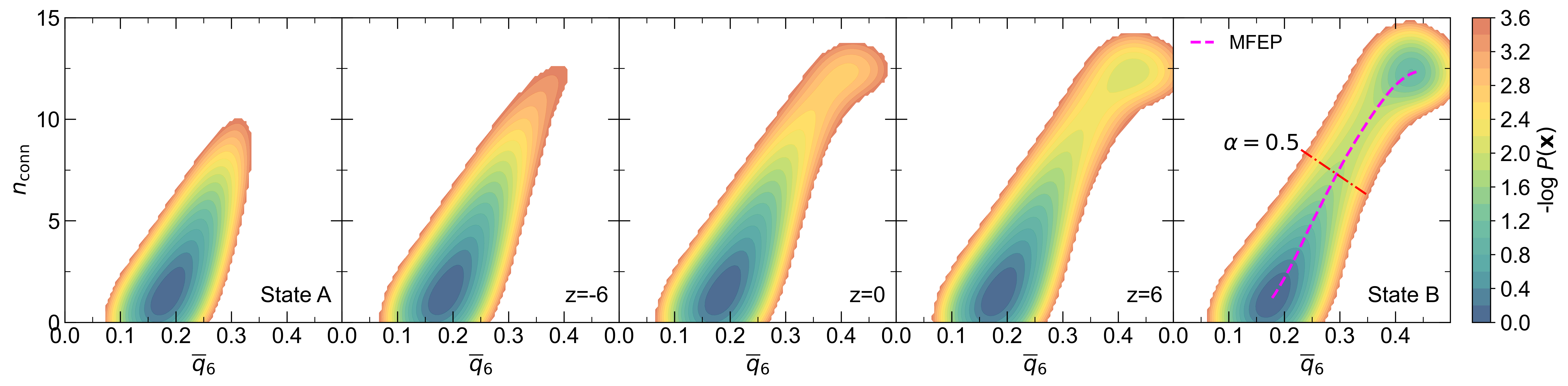}
        \caption{Joint probability distribution of $\overline{q}_6$ and $n_{\text{conn}}$ at different values of $z$. The magenta dashed line shows the highest probability path of the multi-modal distribution. For state $B$ the $\alpha=0.5$ is indicated with a magenta dotted line.}
        \label{fig:3}
    \end{figure*}

\yuanpeng{Simulating first-order phase transitions poses significant technical challenges,\cite{sosso_crystal_2016} and the application of general enhanced sampling methods to these systems requires careful adjustment.
In our case, the high nucleation barrier in units of $ k_\text{B} T $\cite{moroni_interplay_2005, trudu_freezing_2006} will cause numerical issues in optimizing the loss functional for computing $q(\mathbf{x})$.
A helpful numerical improvement here is to replace the variational loss term with $\log \mathcal{K}[q(\mathbf{x})]$ rather than minimizing $\mathcal{K}[q(\mathbf{x})]$ directly\cite{trizio_everything_2025} (see SI Sec.~\ref{sup_sec:committor_training}).}
The two functionals share the same minimum since $\mathcal{K}[q(\textbf{x})]$ is positive definite and the logarithm is a monotonously growing function of its argument. 
\yuanpeng{This approach is crucial for complex systems with high free energy barriers, as it ensures the variational terms in the loss functional will dominate the total loss after the boundary loss is satisfied, allowing it to be optimized effectively. 
Such improved optimization is essential for accurately identifying the TSE in these processes.}

When studying crystallization \yuanpeng{with biased enhanced sampling methods}, it is usually assumed that the initial state $A$ is the liquid and the final state $B$ the fully crystallized simulation box.~\cite{piaggi_enhancing_2017,niu_molecular_2018,niu_temperature_2019}
Since our focus is on studying the TSE and the reactive process, we depart from this procedure and follow the crystallization process until a sufficiently large crystallite is nucleated.
The size of this end crystallite is chosen such that it will irreversibly evolve towards the ordered state \yuanpeng{in accordance with other practices.\cite{moroni_interplay_2005,jungblut_crystallization_2011}}
This approach reduces free energy difference $\Delta G _{A-B}$ and accelerates convergence. 
In such a way, we also reduce the finite size effects that might arise from the size of the growing nucleus and its associated perturbation to the liquid structure being larger than the simulation box, and by the imposition periodic boundary conditions.
The latter implies that the final crystalline configuration has to be commensurate with the simulation box.
If the crystallization axis are not aligned properly and the number of atoms is not correctly chosen, this can lead to defective structures. The formation of such defective structures can be prevented by the use of custom designed \yuanpeng{CVs}, however, this might induce some elements of artificiality.~\cite{karmakar_collective_2021,deng_anisotropic_2023}
In the SI Sec.~\ref{sup_sec:radius_restraint} we illustrate how the $B$ state is generated. 

In order to solve the Kolmogorov variational problem, we express $q(\mathbf {x})$ as a feed forward Neural Network whose features are taken from the vast literature on the problem.~\cite{jung_machine-guided_2023}
In particular, we use as descriptors the Steinhardt parameter ${q}_6^i$,~\cite{steinhardt_bond-orientational_1983} its locally averaged value $\overline{q}_6^i$, and pair entropy $s_S^i$~\cite{piaggi_enhancing_2017,piaggi_entropy_2017} (see SI Sec.~\ref{sup_sec:symbols}).
The rational behind this choice is that we want to describe the balance between enthalpy and entropy that characterize the transition, and these two sets of descriptors can be thought of as proxies these two quantities.
The numerical details of the self-consistent iterative procedure can be also found in the SI Sec.~\ref{sup_sec:committor_training}.

In Fig.~\ref{fig:1} we present the free energy as a function of $z$ and the committor projected along the number of crystal-like atoms in the largest solid cluster $n_\text{c}$ as conventionly done~\cite{rein_ten_wolde_numerical_1996} (see SI Sec.~\ref{sup_sec:definition_nc}).
The marginal of the committor relative to $n_\text{c}$, $q(n_\text{c})$, exhibits the expected step-like behavior, and the critical nucleus estimated from $q(n_\text{c}^*)=0.5$ is $n_\text{c}^*$=220 $\pm$ 20, in agreement with $n_\text{c}^*$ = 240 $\pm$ 34 in Ref.~\cite{trudu_freezing_2006}.
From the free energy plot as a function of $z$ the barrier to nucleation can be estimated to be $\Delta G^*=17.7\pm0.38$ $k_\text{B}T$, where $T=0.725 \, T_m$ is the simulation temperature and $T_m$ is the LJ melting temperature.
This value is in good agreement with previous estimate of $\Delta G^*$ = 18.4 $\pm$ 2 $k_\text{B}T$.~\cite{trudu_freezing_2006}

    \begin{figure}[b]
        \centering
           \includegraphics[width=0.95\linewidth]{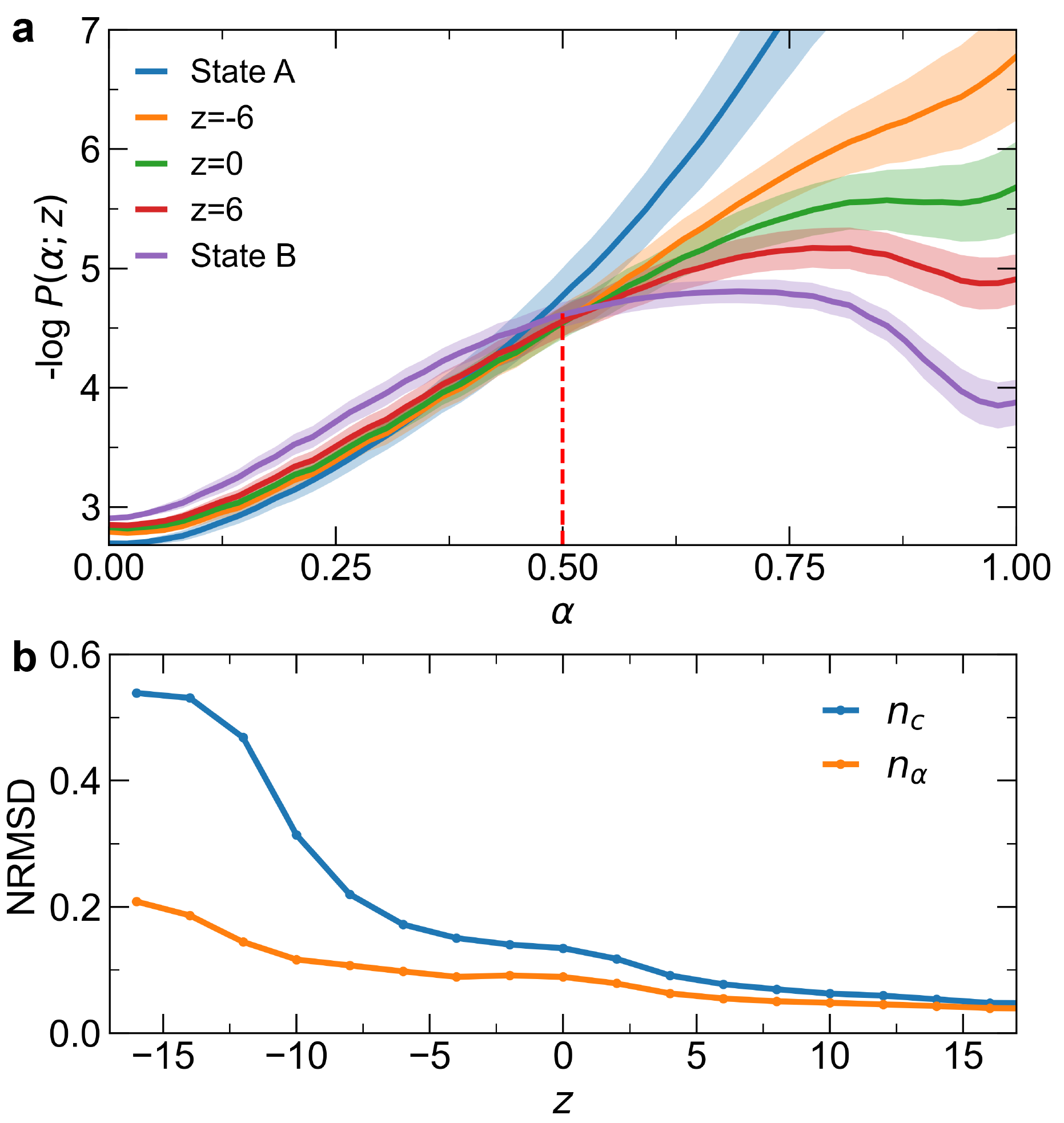}
        \caption{\textbf{a)} Probability distribution $-\text{log}\,P(\alpha;z)$ as function of $\alpha$.
        \textbf{b)} Normalized root mean square deviation (NRMSD) of $n_\text{c}$ and $n_{\alpha}$ along $z$.
        }
        \label{fig:4}
    \end{figure}
    
Having satisfied ourselves of the validity of our approach we use it to better understand the nucleation process.
We start by analyzing the TSE as defined by $p_\mathcal{K}(\mathbf{x})$ in Eq.~\ref{eq:Kolmogorov_probability_pk}, and project our result on the space spanned by the number of crystal-like atoms $n_\text{c}$ and the associated gyration radius $R_g$ \yuanpeng{as shown in Fig.~\ref{fig:2}} (see SI Sec.~\ref{sup_sec:rg_ratio} for the definition of $R_g$).
The TSE thus defined aligns with the minimum free energy path.
In contrast if we were to define the TSE using the standard $q(\mathbf{x}) \simeq 0.5$ criterion, we would have found as TSE a line that is almost orthogonal to minimum free energy path.
Notably, the TSE does not follow the $R_g \propto n_\text{c}^{1/3}$ line, suggesting the critical nuclei are not spherical, \yuanpeng{aligning with experimental results.\cite{moller_crystal_2024}}
\yuanpeng{Our identification of the TSE contains extensive exploration of approximately 3,400 TS configurations in the nucleation process and provides a quantitative assessment of their deviation from ideal spherical geometry, which may offer valuable insights for refining CNT.}
A more detailed analysis of the TSE is provided in the SI Sec.~\ref{sup_sec:analysis_TSE}.

    \begin{figure}[htbp!]
        \centering
           \includegraphics[width=0.95
           \linewidth]{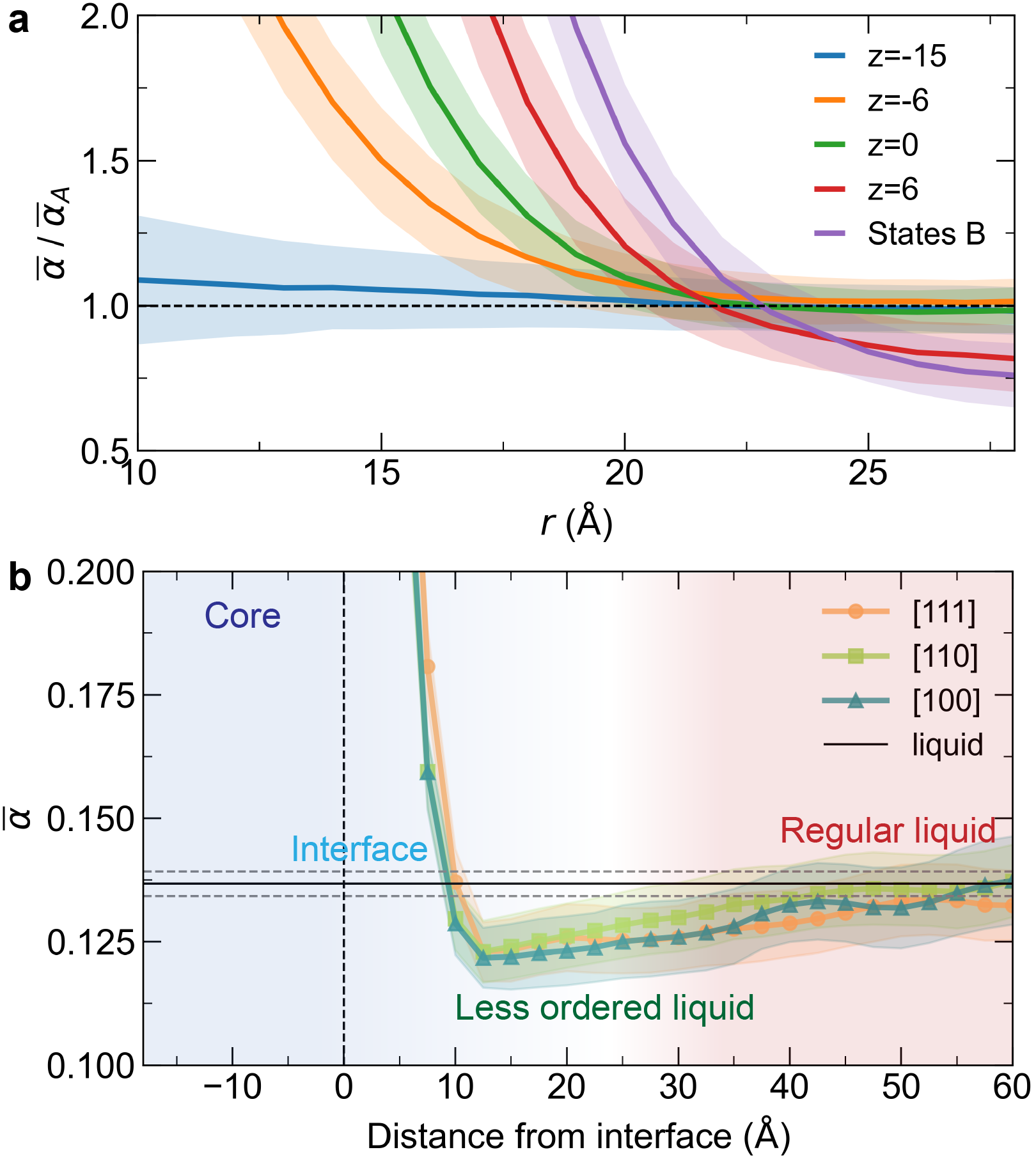}
        \caption{\textbf{a)} Radial distribution function of the averaged per atomic $\alpha^i$, $\overline {\alpha}$, divided by the average value in state $A$, $\overline {\alpha}_{A}$, plotted as a function of distance from the cluster centroid.
        \textbf{b)} Variation of $\overline {\alpha}$ per bin of atoms as a function of distance from the interface, along the [111], [110], and [100] lattice plane.
        Dashed lines represent the standard deviation of pure liquid states.}
        \label{fig:5}
    \end{figure}

We now proceed to study the fluctuations in local liquid order that accompany nucleation. 
\yuanpeng{This is made possible by our probabilistic approach and is facilitated by the fact that the variable $z(\textbf{x})$ induces a natural ordering along the reactive event.}
We shall describe the local order by the joint distribution of \yuanpeng{two bond-order-based parameters $\overline{q}_6$~\cite{steinhardt_bond-orientational_1983} and $n_{\text{conn}}$~\cite{rein_ten_wolde_numerical_1996} as defined in the SI Sec.~\ref{sup_sec:symbols}} and follow how this distribution evolve as a function of $z$ \yuanpeng{(see Fig.~\ref{fig:3})}.
Already in the undercooled liquid state regime, this distribution exhibits a tail towards larger values of $\overline{q}_6$ and $n_{\text{conn}}$. 
As nucleation progresses, this tail grows, and approximately at $z=0$, the distribution becomes multimodal, with a second peak corresponding to the emergence of well-defined growing nuclei.
To simplify the analysis, we project the two-dimensional distribution along the line $\alpha (\overline{q}_6, n_{\text{conn}})$ of highest probability, following a standard procedure used for analyzing multidimensional free energy surfaces~\cite{e_string_2002} (see Fig.~\ref{fig:3}).

As shown in Fig.~\ref{fig:4}\textbf{a}, the $\alpha$ coordinate reveals a clear bimodal distribution distinguishing two phases.
The critical value $\alpha = \frac {1}{2} $ separates the liquid-like regime ($\alpha < \frac {1}{2} $) from the solid-like regime ($\alpha > \frac {1}{2} $).
Thus, we \yuanpeng{estimate the per atomic $\alpha^i$ and} use the $\alpha^i > \frac {1}{2} $ criteria to identify the configurations that belong to the growing crystalline nucleus, \yuanpeng{thereby defining a new probability-based cluster size $n_{\alpha}$ as the number of atoms satisfying this condition.}
The bimodal character of the local order fluctuation suggests that all the atoms in the side peak $n_{\alpha}$ should be counted when measuring the size of the growing nucleus.
The usefulness of this choice is evident in the substantially reduced uncertainty in $z(n_{\alpha})$ compared to $z(n_\text{c})$ (see SI Sec.~\ref{sup_sec:probability_based_nalpha}), as measured by a much reduced normalized root mean square error (see Fig.~\ref{fig:4}\textbf{b}).
Since the fluctuation of $n_{\alpha}$ are clearly smaller than those of $n_\text{c}$, \yuanpeng{it follows that the $n_{\alpha}$ consistently correlated with the reaction event,\cite{moroni_interplay_2005,lam_critical_2023} thus providing a more reliable description of the nucleation process.}

\yuanpeng{When $n_\text{c}$ is used one focuses too much attention on the strictly ordered atoms.
In contrast, the probability-based descriptor $\alpha^i$ enables precise tracking of the evolving degree of order during nucleation.
This approach reveals the radial distribution of $\alpha^i$ from the cluster's center of mass along the reaction coordinate $z$ (see Fig.~\ref{fig:5}\textbf{a}).
Before crossing the TSE, the spatially averaged $\overline \alpha$ decreases smoothly from the nucleus center toward the bulk liquid average $\overline{\alpha}_A$.
After crossing the TSE, the region beyond the nucleus exhibits $\overline{\alpha} / \overline{\alpha}_A < 1$, indicating enhanced disorder relative to the bulk liquid.}

\yuanpeng{To determine whether the observed phenomenon arises from a finite-size simulation box, we further analyzed it in a larger and geometrically simpler case of a planar growing surface as discussed in SI Sec.~\ref{sup_sec:planer_growth}.
Here, the spatially averaged $\overline{\alpha}$ in bins near the solid-liquid interface is significantly lower than $\overline{\alpha}_A$ (see Fig.~\ref{fig:5}\textbf{b}).
Additionally, the bond order parameter $\overline{q}_6$, local entropy, and density also exhibit reduced values near the interface compared to bulk liquid values (see Fig.~\ref{sup_fig:slab_additional}).
This less-ordered region, extending $\sim20 \, \text{\AA}$ from the interface, shows consistent behavior across all tested interface orientations.
Both the cluster and planar approaches provide unified physical insight into the less-ordered liquid region adjacent to the solid-liquid interface.
A plausible explanation of this phenomenon is that, in undercooled liquids, subcritical nuclei continuously form and dissolve.
At the solid-liquid interface, atoms with transient crystal-like order may rapidly incorporated into the growing crystal lattice, resulting in a less-ordered liquid region with reduced order relative to the bulk liquid. 
This finding enriches our understanding of fluctuations in nucleation processes by revealing the existence of a disordered interfacial layer, which may influence the kinetics and thermodynamics of nucleation.
}

\yuanpeng{Finally, we utilize the minimized functional $\mathcal{K}[q(\textbf{x})]$ to estimate the nucleation rate $J$ via the reaction rate $\nu$. 
For an overdamped system with friction coefficient of $\gamma$, at the minimum $\mathcal{K}[q(\textbf{x})]$ is proportional to $\nu$:~\cite{e_transition-path_2010}}

    \begin{equation}
        \nu = \frac{\mathop{ \mathcal{K}[q(\textbf{x})]}}{\beta\gamma} 
        \label{eq:reaction_rate_nu}
    \end{equation}

\noindent \yuanpeng{The effective $\gamma$ is determined by computing the diffusion coefficient $D$ from MD simulations of the undercooled liquid and employing the Einstein relation 
$\gamma = \frac{1}{\beta}\frac{1}{mD}$,\cite{giro_langevin_1985,vogelsang_determination_1987,binder_monte_1995} where $m$ is the mass per particle.}
In such a fashion, we get for $J$ the value \yuanpeng{$\approx 1.25 \times 10^{30} \, \text{m}^{-3} \text{s}^{-1}$} in a much better agreement with the experimental data ($1.5 \text{ - } 6.2 \times 10^{30} \, \text{m}^{-3} \text{s}^{-1}$)~\cite{moller_crystal_2024} than previous theoretical estimation of $3.1 \times 10^{32} \, \text{m}^{-3} \text{s}^{-1}$.~\cite{tipeev_diffusivity_2018}
Details of the estimation of nucleation rate can be found in SI Sec.\ref{sup_sec:nucleation_rate}.


In this work, we found that the method is highly effective in describing nucleation and above all capable of describing in detail the fluctuations that eventually lead to the phase transition. 
The success of our approach opens the way to many developments that will be explored in the future.
We only quote here the possibility of applying a transfer learning approach similar to that of Ref.~\cite{das_correlating_2024} to study with little effort system behavior across different temperatures and across families of similar compounds. 
\yuanpeng{The understanding of fluctuation could lead to better design of crystallization promoting or retarding agents.}

%% file: manuscript/supporting.tex
\setlength{\tabcolsep}{18pt}
\renewcommand{\arraystretch}{1.2}

\section{Generating the $B$ state}
    \label{sup_sec:radius_restraint}
\begin{figure}[h!]
        \centering
        \includegraphics[width=0.4
        \linewidth]{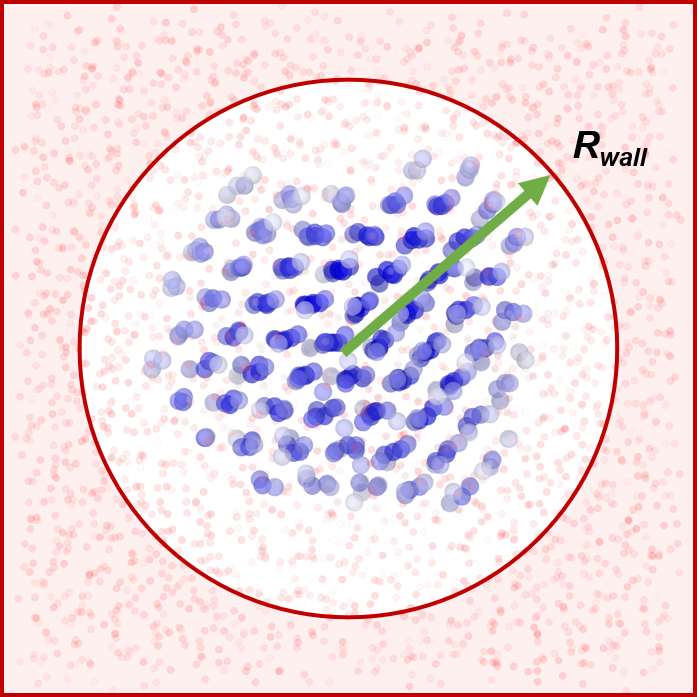}
        \caption{Diagram of the radius restraint setup. Only the atoms in the outer region (shown in red with transparency) contribute to $n_{\overline{q}_6}^\text{out}$, where the bias potential $V_\text{wall}$ is applied. This restraint prevents the inner region from driving complete crystallization of the system.}
        \label{sup_fig:diagram_Rwall}
    \end{figure}

    We implemented a radius restraint as shown in Fig.~\ref{sup_fig:diagram_Rwall}, to accelerate the convergence of OPES simulations and prevent the system from fully crystallizing. 
    First, atoms in the system were identified as solid-like as defined in Eq.~\ref{eq:n_q6}. 
    It should be noted that this $n_{\overline{q}_6}$ differs from the $n_c$ and our probability-based $n_\alpha$, serving merely as a simplified method to distinguish solid-like atoms. 
    An arbitrarily chosen atom was then designated as the reference center of the nucleus, with atoms within a radius $R_\text{wall}$ from this reference atom defined as the inner part, and those beyond it constituting the outer part. 
    A bias potential, $V_\text{wall}$, was subsequently applied to the atoms in the outer part to suppress cluster growth beyond the radius $R_\text{wall}$. 
    The bias potential is expressed as:

    \begin{equation}
        \label{eq:radius_restraint}
        V_\text{wall} =
        \begin{cases} 
        k( n_{\overline{q}_6}^\text{out} - n_\text{th}  )^2,& \text{if } n_{\overline{q}_6}^\text{out} > n_{\text{th}}, \\
        0, & \text{if } n_{\overline{q}_6}^\text{out} \leq n_{\text{th}}.
        \end{cases}
    \end{equation}
    
    where $k = 0.02 \, \text{kJ/mol}$, and a threshold value of $n_{\text{th}} = 200$ was selected to ensure that the bias is activated only when the radius of cluster exceeds $R_\text{wall}$. 
    Furthermore, state $B$ is defined as a sufficiently large crystallite that stabilizes under the influence of $V_\text{wall}$, meaning it will irreversibly transition to the ordered state when simulated with conventional MD. 

\begin{figure}[h!]
        \centering
        \includegraphics[width=0.55
        \linewidth]{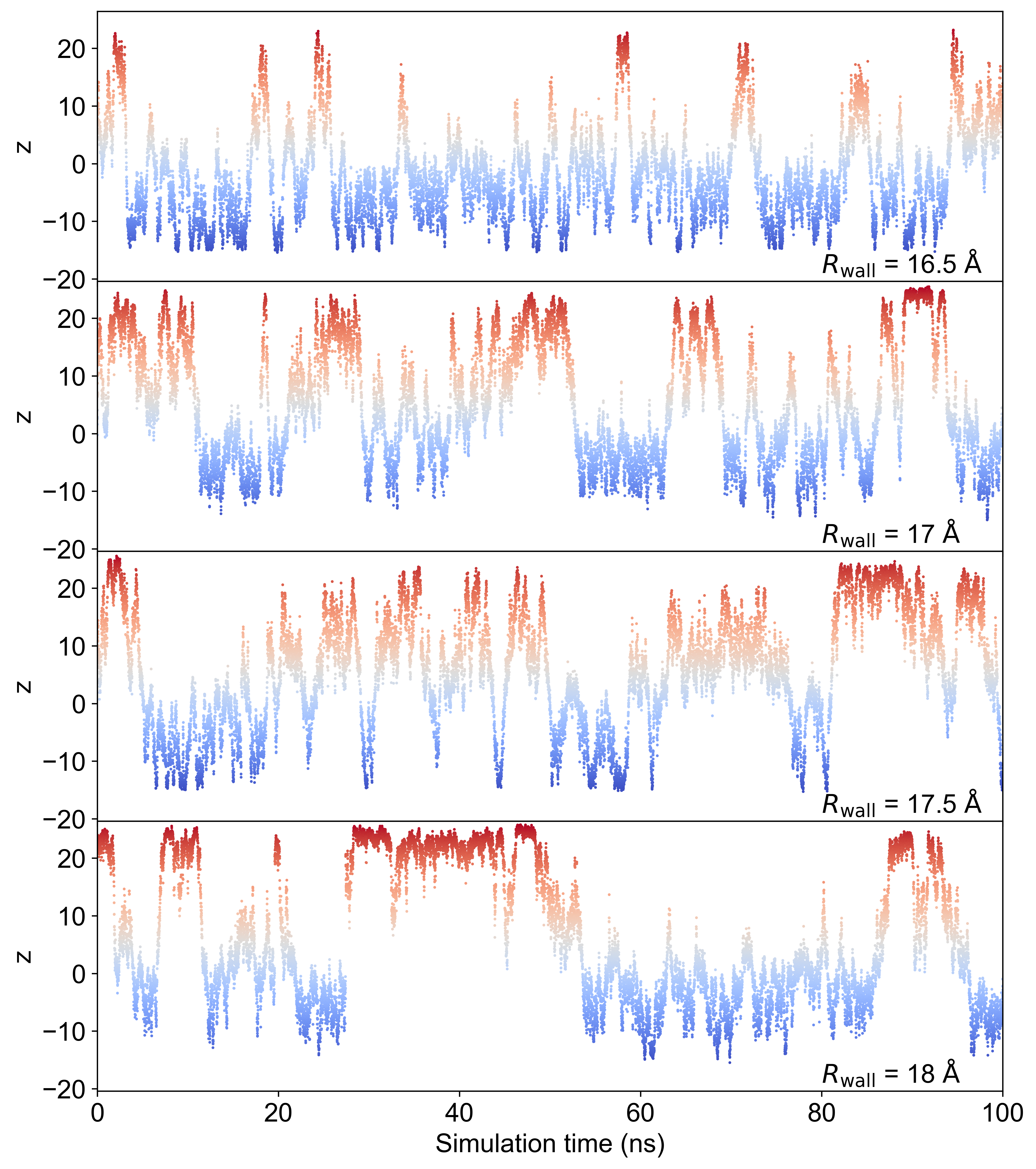}
        \caption{Evolution of $z$ CV as function of time for the 2916-atom LJ system at 67 K with the radius restraint of $R_{\text{wall}}$ = 16.5 $\text{\AA}$, 17 $\text{\AA}$, 17.5 $\text{\AA}$, and 18 $\text{\AA}$.}
        \label{sup_fig:z-t_Rwall}
    \end{figure}

\begin{figure}[h!]
        \centering
        \includegraphics[width=0.5
        \linewidth]{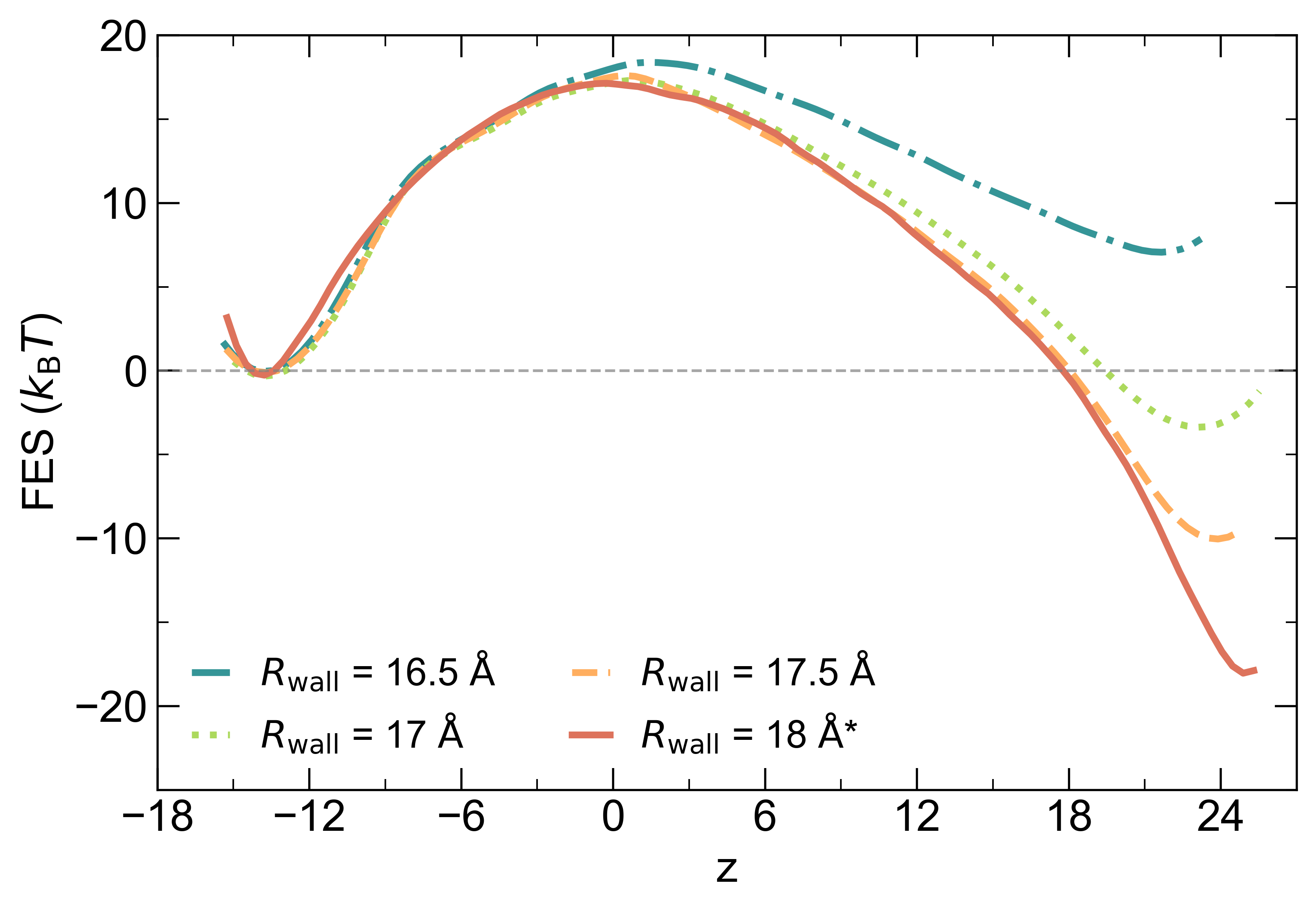}
        \caption{Free energy surface of the 2916-atom LJ system at 67 K with the radius restraint of $R_{\text{wall}}$ = 16.5 $\text{\AA}$, 17 $\text{\AA}$, 17.5 $\text{\AA}$, and 18 $\text{\AA}$.}
        \label{sup_fig:1d_FES_Rwall}
    \end{figure}

    To evaluate the effect of the radius restraint, we performed a series of simulations incorporating OPES with an optimal Kolmogorov bias $V_\mathcal{K}$, and additionally $V_\text{wall}$ with $R_\text{wall}$ values ranging from 16.5 $\text{\AA}$ to 18 $\text{\AA}$. 
    The time evolution of the CV $z$ under different $R_\text{wall}$ conditions is presented in Fig.~\ref{sup_fig:z-t_Rwall}, while the corresponding \yuanpeng{free energy surfaces (FES)} are shown in Fig.~\ref{sup_fig:1d_FES_Rwall}. 
    As shown in Fig.~\ref{sup_fig:1d_FES_Rwall}, the FES exhibits gradual convergence with increasing $R_\text{wall}$.
    With a radius restraint of $R_\text{wall} = 18 \,\text{\AA}$, the FES in the region where $z < 18$ closely matches that obtained at $R_\text{wall} = 17.5 \,\text{\AA}$. 
    As a result, we adopted $R_\text{wall} = 18 \,\text{\AA}$ for all subsequent analyses, ensuring that the inner part encompasses the size of the critical nuclei while minimizing boundary effects. 
    All the statistical analyses were conducted within the region $z < 18$ which remains unaffected by the bias of radius restraint.
    To further ensure accuracy, configurations with $V_\text{wall} > 0$ were excluded from the statistical calculations.

\clearpage
\newpage

\section{Symbols and their definition}
    \label{sup_sec:symbols}

\subsection{Steinhardt bond order descriptors}
    We selected a set of Steinhardt bond order descriptors~\cite{steinhardt_bond-orientational_1983,rein_ten_wolde_numerical_1996,lechner_accurate_2008} to characterize atomic local environment for LJ system.
    First, the complex vector of each atom $i$, $q_{lm}^i$, is defined as:
    \begin{equation}
        q_{lm}^i = \frac{\sum_j \sigma(r_{ij}) Y_{lm}(\textbf(r_{ij}))}{\sum_j \sigma(r_{ij})}.
        \label{eq:qlmi}
    \end{equation}
    
    \noindent where $Y_{lm}$ are spherical harmonics with degree of $l$, vector $\textbf(r_{ij})$ connects atom $i$ and $j$, and $N_b$ is the number of neighbors of atom $i$.
    A switching function is used to make the function decays smoothly to 0 at the cutoff of first coordinate shell,
    
    \begin{equation}
        \label{eq:switching_function1}
        \sigma(r_{ij}) =
        \begin{cases} 
        {\color{black} \frac{1 }{1 + ( r_{ij}/r_c  )^{6}} }, & \text{if } r_{ij}<r_{\text{max}}, \\
        0, & \text{if } r_{ij} \geq r_{\text{max}}.
        \end{cases}
    \end{equation}
    
    \noindent where \yuanpeng{$r_c=4.5\,\text{\AA}\approx 1.32\,\sigma$} is set to ensure the contribution reduced smoothly at \yuanpeng{$r_{\text{max}}=5.1\,\text{\AA}\approx 1.5\,\sigma$}, corresponding to the first minimum of the radial distribution function of the liquid.
    Then, the $l$ order Steinhardt parameter $q_l^i$ is given as:

    \begin{equation}
        q_l^i = \sqrt{\frac{4\pi}{2l + 1} \sum_{m=-l}^{l} |q_{lm}^i|^2}.
    \end{equation}

    To improve its performance, we can replace the $q_{lm}$ vectors of atom $i$ by the average over its nearest neighbors and itself with the same switching function as Eq.~\ref{eq:switching_function1},

    \begin{equation}
        \overline{q}_{lm}^i = \frac{\sum_j \sigma(r_{ij}) \sum_{m=-6}^{6} q_{6m}^{i\,\,*} q_{6m}^j}{\sum_j \sigma(r_{ij})}.
            \label{eq:local_qlm}
    \end{equation}

    \noindent Then, the local averaged Steinhardt parameter, $\overline{q}_l^i$, can be given as:

    \begin{equation}
        \overline{q}_l^i = \sqrt{\frac{4\pi}{2l + 1} \sum_{m=-l}^{l} |\bar{q}_{lm}^i|^2},
    \end{equation}

    While $q_l^i$ holds the structure information of the first coordinate shell around atom $i$, its averaged form $\overline{q}_l^i$ incorporates the second shell, offering a more comprehensive description of the local environment.
    The histogram of $q_6^i$ and $\overline{q}_6^i$ are estimated using a Gaussian expansion to ensure continuous derivatives and are presented in Fig.~\ref{sup_fig:descriptor_distribution}\textbf{a} and \textbf{b}, respectively.

    Given the critical role of the number of solid-like atoms in classical nucleation theory, we define $n_{\overline{q}_6}$ as a smooth descriptor approximating the count of atoms with $\overline{q}_6^i > 0.325$, achieved by combining it with a switching function. 
    This provides a simplified yet effective measure of the number of solid-like atoms, expressed as: 

    \begin{equation}
        \label{eq:n_q6}
        n_{\overline{q}_6} = \sum_i {\color{black} \frac{1 }{1 +( \overline{q}_6^i/0.325 )^{6}} }
    \end{equation}

    For the Steinhardt bond-order parameters, we utilize a total of five bins from the histograms of $q_6^i$ and $\overline{q}_6^i$, along with the averaged $\overline{q}_6$ and $n_{\overline{q}_6}^{\text{in}}$ as descriptors.

\clearpage
\newpage 

\subsection{Pair Entropy}
    To better capture longer range atomic environment, we take the entropy-based fingerprint~\cite{piaggi_entropy_2017} to further identify local structures and used as descriptors for the training of committor model.
    The projection of an approximate evaluation of the entropy on atom $i$ can be expressed as:

    \begin{equation}
        s_S^i = -2\pi \rho k_B \int_{0}^{r_m} [ g_m^i(r) \ln g_m^i(r) - g_m^i(r) + 1 ] r^2 dr,
        \label{eq:pair_entropy}
    \end{equation}

    \noindent where \yuanpeng{$r_m = 7.5\,\text{\AA}\approx 2.2\,\sigma$} is an upper integration limit , and $g_m^i(r)$ is the radial distribution function centered at the atom $i$ given as:

    \begin{equation}
        g_m^i(r) = \frac{1}{4\pi \rho r^2} \sum_{j} \frac{1}{\sqrt{2\pi \sigma^2}} e^{-(r - r_{ij})^2 / (2\sigma^2)},
        \label{eq:rdf_entropy}
    \end{equation}

    \noindent where $j$ are the neighbors of atom $i$, $r_{ij}$ is the distance between atoms $i$ and $j$, and $\sigma = 0.5 \,\text{\AA}$ is a broadening parameter.
    \yuanpeng{Furthermore, the average local entropy can be defined as,}

    \begin{equation}
        \overline{s}_{S}^i = \frac{\sum_j \sigma(r_{ij}) {s}_{S}^j+{s}_{S}^i}{\sum_j \sigma(r_{ij})+1},
        \label{eq:local_entropy}
    \end{equation}

    \noindent \yuanpeng{where the $\sigma(r_{ij})$ take the same setting as in Eq.\ref{eq:switching_function1}.}

\begin{figure}[h!]
        \centering
        \includegraphics[width=1
        \linewidth]{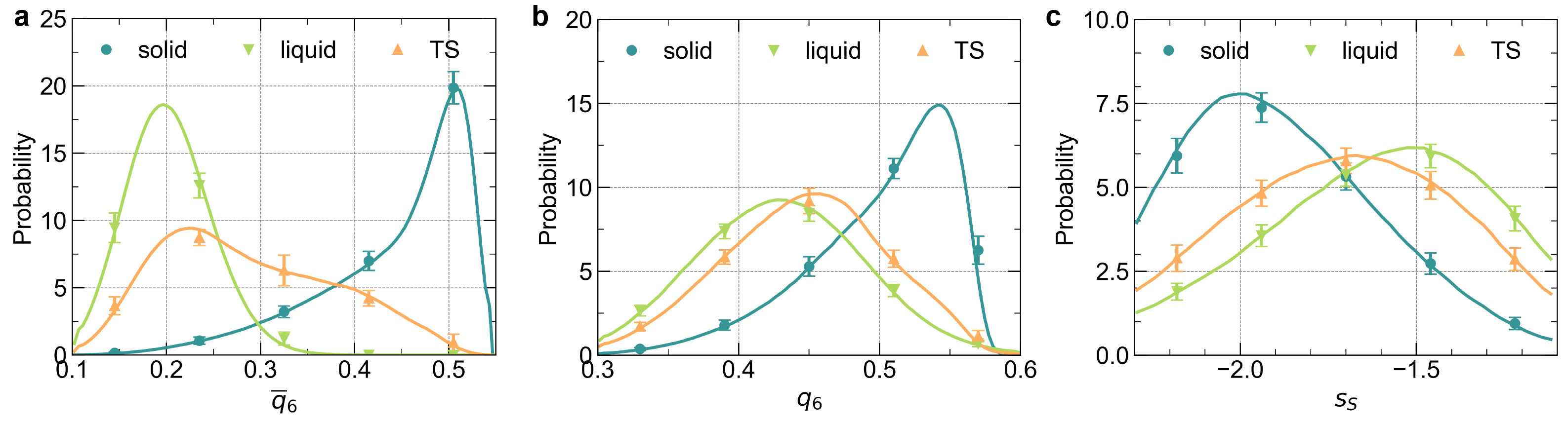}
        \caption{Histogram of \textbf{a)} $\overline{q}_6^i$,\textbf{b)} ${q}_6^i$, and \textbf{c)} $s_S^i$.
        Dotted point refer to the average of 5 bins selected as the descriptors.}
        \label{sup_fig:descriptor_distribution}
    \end{figure}

    The histogram of $s_S^i$ is estimated using a Gaussian expansion and is shown in Fig.~\ref{sup_fig:descriptor_distribution}\textbf{c}.
    The values of five bins from the histograms $s_S^i$ are selected as descriptors.
    \yuanpeng{The locally averaged $\overline{s}_{S}^i$ is used for further estimating the atomic environment near the solid-liquid interface as discussed in Sec.\ref{sup_sec:planer_growth}.}

\subsection{Counting the number of crystal-like atoms in the growing nuclei; $n_\text{c}$}
    \label{sup_sec:definition_nc}

    We first identified crystal-like atoms within the largest cluster, $n_{\text{c}}$, based on the criterion introduced in Ref.~\cite{rein_ten_wolde_numerical_1996}.
    Two particles are classified as neighbors if their distance is less than 5 $\text{\AA}$, \yuanpeng{regarding to 1.47 $\sigma$}.
    The normalized scalar product of the complex vectors $q_{6m}$ for neighboring particles $i$ and $j$ is then given by:

    \begin{equation}
        S_{ij} = \frac{\sum_{m=-6}^{6} q_{6m}(i) q_{6m}^*(j)}{ ( \sum_{m=-6}^{6} |q_{6m}(i)|^2  )^{1/2}  ( \sum_{m=-6}^{6} |q_{6m}(j)|^2  )^{1/2}},
    \end{equation}

    \noindent which quantifies the degree of structural correlation between the local environments of particles $i$ and $j$.
    If this scalar product is larger than 0.5, the two particles are considered to be connected to each other.
    \yuanpeng{The number of connections per particle is defined as a per atom quantity $n_{\text{conn}}$.}
    A particle is identified as crystal-like, if its number of connections $n_{\text{conn}}$ is larger than 8.
    Crystal-like particles that are neighbors belong to the same cluster.
    The size of the largest cluster is denoted as $n_{\text{c}}$.

\clearpage
\newpage 

\subsection{Gyration radius, shape descriptor, and surface area ratio}
    \label{sup_sec:rg_ratio}
    
    To gain deeper insight into the evolution of cluster shapes during the nucleation process, \yuanpeng{we estimated the gyration tensor for each cluster and analyzed its relative gyration radius and shape anisotropy $\kappa^2$.~\cite{arkin_gyration_2013}
    First, we determined the center of mass of the cluster $\mathbf{r}_{\text{com}}$.}
    The gyration matrix $\underline{R}_g$ of a cluster is then computed as:
    
    \begin{equation}
        \underline{R}_g^2 = \frac{1}{N} \sum_{i=1}^{N} (\mathbf{r}_i - \mathbf{r}_{\text{com}}) \otimes (\mathbf{r}_i - \mathbf{r}_{\text{com}}),
    \end{equation}

    \noindent where $N$ is the number of atoms forming the cluster, $\mathbf{r}_i$ is the position of the particles belonging to the cluster.

    \yuanpeng{The eigenvalues of the tensor $\underline{R}_g^2$, denoted as $\lambda _1$, $\lambda _2$, and $\lambda _3$ are sorted in ascending order, \textit{i.e.}, $\lambda _1 \leq \lambda _2 \leq \lambda _3$.}
    The gyration radius $R_g$ of the cluster is then given as:

    \begin{equation}
        {\color{black}R_g = [ \langle \lambda _1 \rangle + \langle \lambda _2 \rangle + \langle \lambda _3 \rangle ]^{1/2}.}
    \end{equation}

   \yuanpeng{To characterize the shape of the cluster, we calculated the second invariant shape descriptor $\kappa^2$:~\cite{arkin_gyration_2013}}

    \begin{equation}
        {\color{black}\kappa^2 = 1 - 3 \frac{\lambda_1 \lambda_2 + \lambda_2 \lambda_3 + \lambda_3 \lambda_1}{(\lambda_1 + \lambda_2 + \lambda_3)^2}.}
    \end{equation}

    \yuanpeng{This parameter ranges from 0 for spherically symmetric clusters to 1 for highly anisotropic clusters, enabling a comprehensive analysis of the cluster's shape evolution during nucleation.}
    
    \yuanpeng{It should be notice that the $\kappa^2$ may fail to distinguish highly symmetric shapes such as tetrahedra from spheres.
    We further take the surface area ratio ${A_{\text{cluster}}}/{A_{\text{sphere}}}$, as defined by the ratio of area of cluster surface to that of the area of the surface of the ideal sphere of the same volume as the cluster, to provide a more direct measure of shape deviation from sphericity.
    The surface area ratio is dimensionless, approaching 1 for spherical clusters and exceeding 1 for non-spherical shapes.
    To start with, we identified atoms in the largest crystalline cluster with both the conventional cluster size $n_\text{c}$ and our probability-based $n_\alpha$ as criteria.
    Then, the $A_{\text{cluster}}$ was computed as the surface area of the convex hull\cite{barber_quickhull_1996} of cluster by using SciPy.\cite{virtanen_scipy_2020}
    For the surface area of an ideal sphere $A_{\text{sphere}}$, its radius $R$ can be obtained as,}
    
    \begin{equation}
        {\color{black}R=(\frac{3}{4}\frac{n}{\pi\rho_s})^{1/3}.}
    \end{equation}

    \noindent \yuanpeng{where $n$ is the number of atoms in the largest crystalline cluster, and $\rho_s$ is the number density of the solid phase.
    Detailed results of the evolution of shape descriptor $\kappa^2$ and surface area ratio ${A_{\text{cluster}}}/{A_{\text{sphere}}}$ are discussed in Sec.~\ref{sup_sec:analysis_TSE}.}

\clearpage
\newpage

\section{Committor model training}
    \label{sup_sec:committor_training}

\subsection{Learning the committor function}
    We follow the similar self-consistent procedure described in Ref.~\cite{kang_computing_2024,trizio_everything_2025} for solving the \yuanpeng{committor} function. 
    The committor function is represented as a \yuanpeng{neural network (NN)} $q_\theta(\mathbf{x}) = q(\mathbf{d}(\mathbf{x}))$ with learnable parameters $\theta$ that takes as inputs a set of physical descriptors $\mathbf{d}(\mathbf{x})$.
    The optimization of $q_\theta^n(\mathbf{x})$, at iteration $n$, is based on the minimization of an objective function $\mathcal{L}$ that formalizes the variational principle of Eq.~\ref{eq:variational_functional}, $\mathcal{L}_{v}$, and its boundary conditions $\mathcal{L}_{b}$. 
    The relatively high nucleation barrier at temperature below $T_m$~\cite{moroni_interplay_2005,trudu_freezing_2006} may led to significant imbalance between $\mathcal{L}_{v}$ and $\mathcal{L}_{b}$.
    Specifically, the variational loss was several orders of magnitude smaller than the boundary loss, resulting in suboptimal performance.

    To address this issue, here we modified an optimized logarithmic loss function as,

    \begin{equation}
        \mathcal{L} = \log \mathcal{L}_{v} + \alpha \mathcal{L}_{b}
        \label{eq:loss_function_new}
    \end{equation}

    \noindent which is relatively scaled by the hyperparameter $\alpha$. The $L_v$ term derives from the variational principle and reads:

    \begin{equation}
        \mathcal{L}_v = \frac{1}{N_n} \sum_{i=1}^{N_n} w_i |\nabla_\mathbf{u} q(\mathbf{x}_i)|^2,
        \label{eq:loss_variational}
    \end{equation}

    \noindent where the weights $w_i$ weigh each configuration $\mathbf{x}_i$ to natural Boltzmann weight.
    The $L_b$ term enforces the boundary conditions $q(\mathbf{x}_A) = 0$ and $q(\mathbf{x}_B) = 1$ as:
    
    \begin{equation}
        \mathcal{L}_b = \frac{1}{N_A} \sum_{i \in A} (q(\mathbf{x}_i) - 0)^2 + \frac{1}{N_B} \sum_{i \in B} (q(\mathbf{x}_i) - 1)^2.
        \label{eq:loss_boundary}
    \end{equation}

    As discussed in Ref.~\cite{trizio_everything_2025}, the $\mathcal{L}_b$ term was evaluated using a labeled dataset of $N_A$ and $N_B$ configurations collected at the first iteration via unbiased simulations. 
    In contrast, the $\mathcal{L}_v$ term at the first iteration was solely influenced by these unbiased labeled configurations, whereas from the second iteration onward, only the biased configurations contributed to the $\mathcal{L}_v$ term. 
    This strategy can help get rid of the slowdown arising from the incorrect weights of the unbiased configurations in the dataset.

\subsection{Self-consistent iterative procedure}
    In Ref.~\cite{kang_computing_2024}, we demonstrated that the variational principle, combined with the sampling of the Kolmogorov ensemble, can be used in a self-consistent iterative framework to obtain the committor function with the assistance of machine learning.
    This iterative procedure alternates between training a NN-based variational parametrization of the committor function and generating progressively improved data through sampling under the influence of the TS-oriented Kolmogorov bias $V_\mathcal{K}(\textbf{x})$.
    In this work, the procedure begins with statistical sampling driven by $V_\mathcal{K}^n(\textbf{x})$ to estimate the functional $\mathcal{K}[q(\textbf{x})]$ and obtain an initial approximation of the committor $q^0(\textbf{x})$.
    Subsequently, $q(\textbf{x})$ is refined iteratively using a combination of the Kolmogorov bias $V_\mathcal{K}^n(\textbf{x})$ and an adaptive bias $V_{\text{OPES}}(\textbf{x})$\cite{trizio_everything_2025}.
    A schematic overview of the iterative steps is provided as below.


    \begin{itemize}
        \item \textbf{Step 1:} An initial approximation of the committor $q^0(\mathbf{x})$ is obtained through an iterative procedure involving the Kolmogorov bias potential $V_\mathcal{K}$. At the first iteration ($n = 0$), a dataset is generated by performing short unbiased simulations within the metastable basins, and the sampled configurations are labeled accordingly. The initial committor model $q^0(\mathbf{x})$ is then trained using configurations $\mathbf{x}^n$ and weights $w_i^n$ obtained under $V_\mathcal{K}$, following the variational principle defined in Eq.~\ref{eq:loss_function_new}.

        \item \textbf{Step 2:} Biased simulations are carried out under the combined influence of the Kolmogorov bias potential $V_\mathcal{K}^n$ and the OPES bias potential $V_\text{OPES}^n$.
        The OPES bias acts along the committor-based collective variable $z^n(\mathbf{x})$, promoting efficient sampling across the configurational space between metastable states.
        Meanwhile, the Kolmogorov bias ensures thorough sampling of the TS region, which is essential for refining the committor model via the variational principle.

        \item \textbf{Step 3:} The sampled configurations are added to the dataset and reweighted using the total effective bias potential $V_\text{eff} = V_\mathcal{K} + V_\text{OPES}$.
        The reweighting factor is given by $w_i^n = \frac{e^{\beta V_\text{eff}^n(\mathbf{x}_i)}}{\langle e^{\beta V_\text{eff}^n(\mathbf{x})} \rangle_{U_\text{eff}^n}}$, where $\beta$ is the inverse temperature.
        This reweighting step ensures a progressively accurate estimate of the FES as the iterations proceed.

        \item \textbf{Step 4:} The updated dataset, containing reweighted configurations and their associated weights $w_i^n$, is used to re-optimize the neural-network-based committor $q^n(\mathbf{x})$ using the variational principle of Eq.~\ref{eq:loss_function_new}. The refined committor model serves as input for the next iteration from step 2. This iterative process progressively improves the accuracy of both the committor and the estimated FES.
    \end{itemize}

\subsection{Training parameters}
    To model the committor function $q_\theta(\mathbf{x})$ at each iteration, we used $n_{\overline{q}_6}^{\text{in}}$, $\overline{q}_6$, along with the histograms of $q_6^i$, $\overline{q}_6^i$, and $s_S^i$ as inputs to a NN with the architecture [17, 80, 40, 1].
    The optimization was performed using the ADAM optimizer with an initial learning rate of $10^{-3}$, modulated by an exponential decay with a multiplicative factor of $\gamma = 0.99993$.
    The training ran for approximately $\sim$50,000 epochs, with the $\alpha$ hyperparameter in the loss function Eq.~\ref{eq:loss_function_new} set to 5.
    The $K_m$ values of the $q_\theta(\mathbf{x})$ model trained with different descriptor sets at final iteration are presented in Table~\ref{sup_tab:performance_descriptor}.
    The number of iterations, corresponding dataset sizes, $\lambda$ values, and OPES \texttt{BARRIER} parameters used in the biased simulations are summarized in Tables~\ref{sup_tab:training_iteration_Vk} and ~\ref{sup_tab:training_iteration_Vk_OPES}.
    Additionally, these tables report the lowest values of the functional $K_m$ as a measure of model quality and convergence, along with the simulation time $t_s$ and the output sampling time $t_o$.

\subsection{OPES+$V_\mathcal{K}$ biased simulations}
    \label{sup_sec:setting_biased_simulations}
    We employed OPES enhanced sampling techniques to investigate the reversible nucleation process of the LJ system.
    The committor value $z$ was used as the CV to drive the OPES simulations.

    In OPES method, the equilibrium probability distribution $P(z)$ is estimated on the fly, and a bias potential $V_n(s)$ is constructed to drive $P(z)$ toward a target distribution $P_\text{tg}(z)$.
    Here, we chose the well-tempered distribution~\cite{bonomi_reconstructing_2009} as the target:

    \begin{equation}
        P^{\text{tg}}(s) \propto \left[ P(s) \right]^{\frac{1}{\gamma}},
    \end{equation}

    \noindent where $\gamma > 1$ is the bias factor.
    In this framework, the bias potential at the $n$-th iteration is expressed as:

    \begin{equation}
        V_n(s) = \left( 1 - \frac{1}{\gamma} \right) \frac{1}{\beta} \log \left( \frac{P_n(s)}{Z_n} + \varepsilon \right),
    \end{equation}

    \noindent where $Z_n$ is a normalization factor, $\beta$ is the inverse temperature, and $\varepsilon = e^{-\beta \Delta E / ( 1 - \frac{1}{\gamma} )}$ is a regularization parameter that limits the maximum deposited bias $\Delta G$.
    In this study, we set the kernel deposition frequency to 500 and restricted $\Delta G$ to 15~kJ~mol$^{-1}$ (corresponding to the \texttt{STRIDE} and \texttt{BARRIER} parameters in the PLUMED input files, respectively).

\begin{figure}[h!]
        \centering
        \includegraphics[width=0.5
        \linewidth]{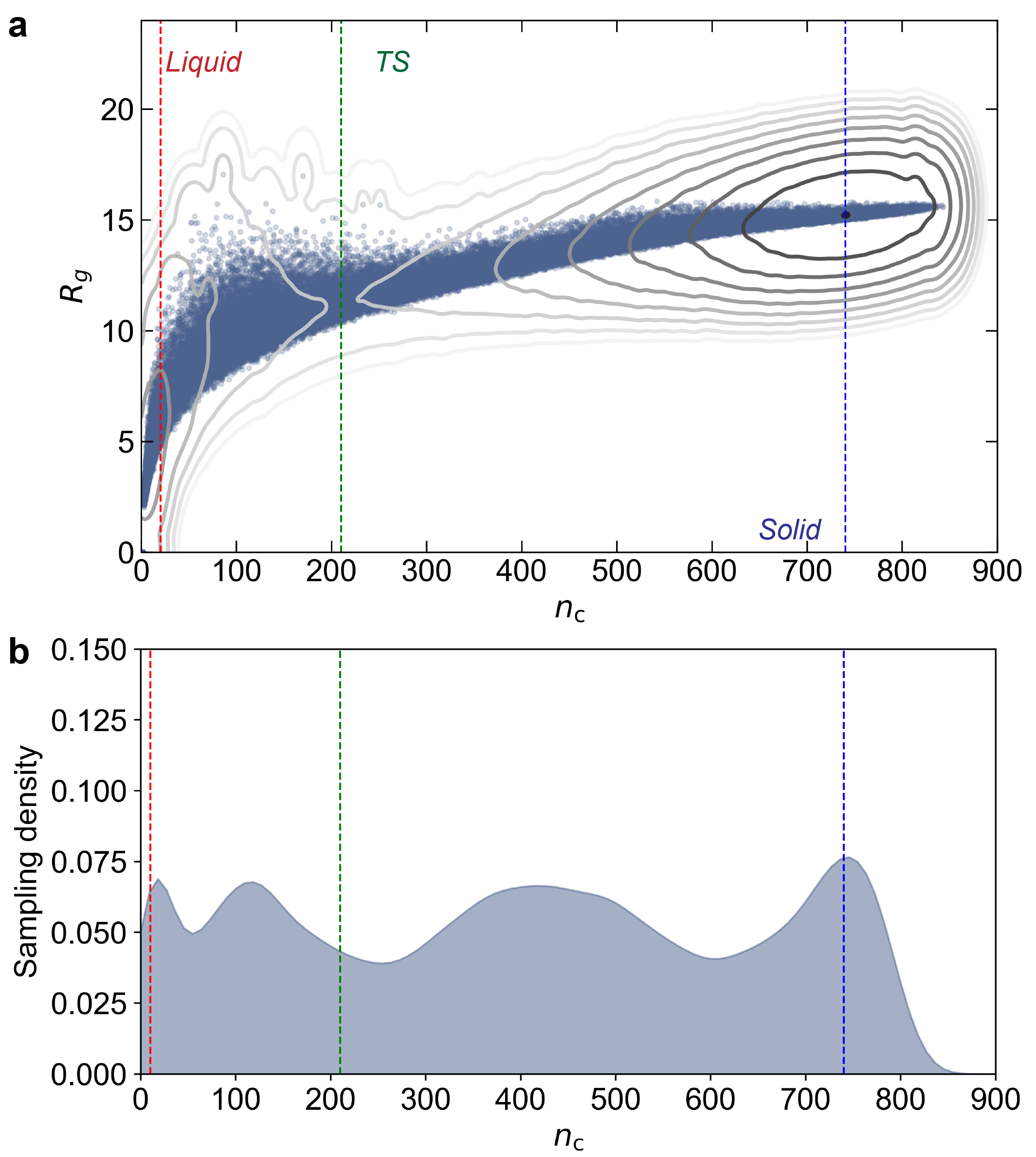}
        \caption{\textbf{a)} Scatter plot of the sampled points under the bias of OPES+$V_\mathcal{K}$.
        In the background, the FES is represented by its isolines.
        \textbf{b)} Distribution of the sampled configurations along $n_{\text{c}}$ in the simulation. 
        This picture demonstrates that the sampling is uniform in both metastable and TS region.}
        \label{sup_fig:sampling_density}
    \end{figure}

\begin{figure}[h!]
        \centering
        \includegraphics[width=0.5
        \linewidth]{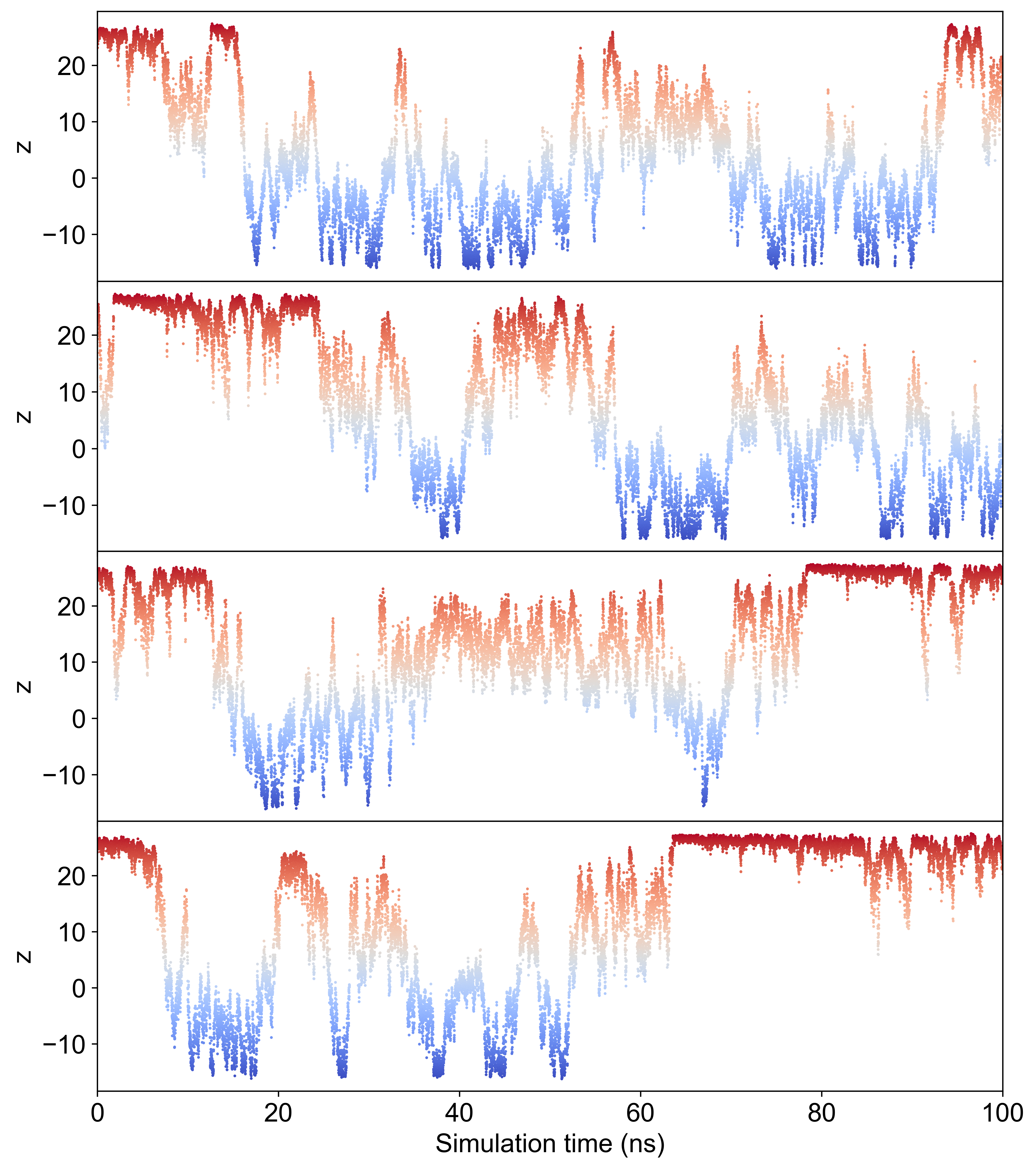}
        \caption{Evolution of $z$ CV as function of time for the 2916-atom LJ system at 67 K. Four independent simulations with different random seeds are shown, demonstrating the efficiency of our sampling method.}
        \label{sup_fig:z-t}
    \end{figure}

\begin{figure}[h!]
        \centering
        \includegraphics[width=0.5
        \linewidth]{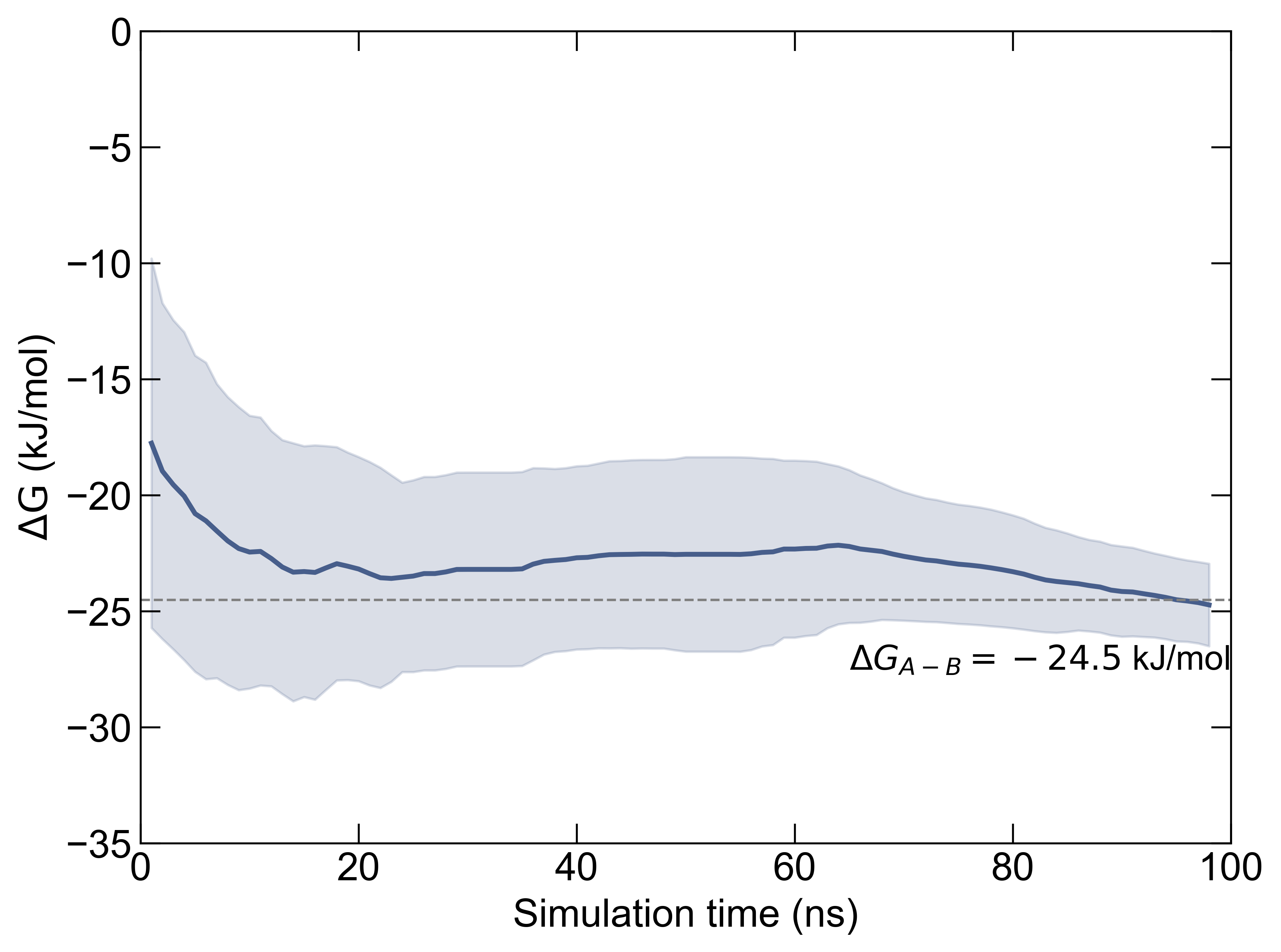}
        \caption{Convergence with simulation time of the estimate $\Delta G_{A-B}$ over four set of OPES+$V_\mathcal{K}$ simulations for the 2916-atom LJ system at 67 K.}
        \label{sup_fig:free_energy_difference}
    \end{figure}

    The sampling density and the time evolution of the $z$ CV are shown in Fig.~\ref{sup_fig:sampling_density} and Fig.~\ref{sup_fig:z-t}, respectively.
    The convergence of the solid-liquid free energy difference $\Delta G_{A-B}$ is presented in Fig.~\ref{sup_fig:free_energy_difference}.

    For the sampling of dataset, we used LJ parameter for argon, with $\epsilon$ = 119.8 K and $\sigma$ = 0.3405 nm.~\cite{rowley_monte_1975}
    The interaction potential cutoff is set at \yuanpeng{9 $ \text{\AA} \,\approx 2.64\,\sigma$} and long-range corrections are considered in the computation of both total energy and pressure.~\cite{sun_compass_1998}
    A timestep of 10 fs, \yuanpeng{regarding to $\approx \,0.0047 \tau$} was used for the \yuanpeng{integration} of motion in biased simulations.
    The Nosé\text{-}Hoover thermostat~\cite{hoover_canonical_1985} and Parrinello\text{-}Rahman barostat~\cite{parrinello_polymorphic_1981} were used with a relaxation time of 0.1 ps and 1 ps, respectively.
    Based on the regression function of melting line in Ref.~\cite{mastny_melting_2007}, we take T = 67 K referred to 0.725 $T_m$ at P = 0.25 kbar for all the simulations in current work.

\newpage

    \begin {table}[h!]
        \caption {Comparison of the performances of different descriptor sets for LJ system. The descriptor sets discussed in the main text are marked with an asterisk.} 
        \label{sup_tab:performance_descriptor}
        \begin{center}
        \begin{tabular}{ |c|c|c|c|c|c| } 
         \hline
         Set id & \yuanpeng{Descriptors $d(\textbf{x})$} & $K_m$ \\ 
         \hline
             1 & $n_{\overline{q}_6}^{\text{in}}$, $\overline{q}_6$ & 3.08 \\
             2 & histogram of $\overline{q}_6^i$ & 3.68 \\
             3 & Set 1 + Set 2 & 2.82 \\
             4 & Set 1 + histogram of $q_6^i$ and $s_S^i$ & 2.81 \\
             5* & Set 2 + Set 4 & 2.54 \\
         \hline
        \end{tabular}
        \end{center}
    \end {table}

    \begin {table}[h!]
        \caption {Summary of the initial estimation with $V_\mathcal{K}$ biased simulations for LJ system.} 
        \label{sup_tab:training_iteration_Vk}
        \begin{center}
        \begin{tabular}{ |c|c|c|c|c|c| } 
         \hline
         Iteration & Dataset size & $K_m$ & $\lambda$ & $t_s$ [ns] & $t_o$ [ps] \\ 
         \hline
            0  & 2000 &  $9.5 \times 10^{11}$ &  -   & 2*0.5 & 0.5\\
            1  & 6000 &  $7.3 \times 10^{4}$  &  3   & 2*0.4 & 0.2\\
            2  & 6000 &  $9.5 \times 10^{3}$  &  2   & 2*0.4 & 0.2\\
            3  & 10000 & $6.2 \times 10^{3}$  &  1.2-1.35  & 4*0.4 & 0.2\\
         \hline
        \end{tabular}
        \end{center}
    \end {table}

    \begin {table}[h!]
        \caption {Summary of the iterative procedure with OPES+$V_\mathcal{K}$ biased simulations for LJ system.} 
        \label{sup_tab:training_iteration_Vk_OPES}
        \begin{center}
        \begin{tabular}{ |c|c|c|c|c|c| } 
         \hline
         Iteration & Dataset size & $K_m$ [au] & OPES \texttt{BARRIER} [kJ/mol] & $t_s$ [ns] & $t_o$ [ps] \\ 
         \hline
            1   & 22000 &  835.5 & 15 & 4*4 & 0.4\\
            2   & 22000 &  144.7 & 15 & 4*10 & 1\\
            3   & 22000 &  3.43 & 15 & 2*50 & 5\\
            4   & 22000 &  4.11 & 15 & 4*50 & 10\\
            5   & 22000 &  2.54 & 15 & 4*100 & 20\\
         \hline
        \end{tabular}
        \end{center}
    \end {table}

\clearpage
\newpage

\clearpage
\newpage

\section{Analysis of TSE}
    \label{sup_sec:analysis_TSE}

    \yuanpeng{To investigate the shape evolution and aspherical nature of critical nuclei, we analyzed the TSE defined by $p_\mathcal{K}(\mathbf{x})$ in Eq.~\ref{eq:Kolmogorov_probability_pk}.
    We projected the TSE onto two-dimensional spaces defined by combinations of the cluster sizes ($n_\text{c}$ or $n_\alpha$) and the shape descriptors $\kappa^2$ and the surface area ratio $A_{\text{cluster}}/A_{\text{sphere}}$.
    Due to challenges in accurately estimating the interface area for very small crystal clusters using the convex hull method, we restricted our statistical analysis with surface area ratio to clusters with $n > 50$.}
    
\begin{figure}[h!]
        \centering
        \includegraphics[width=0.75
        \linewidth]{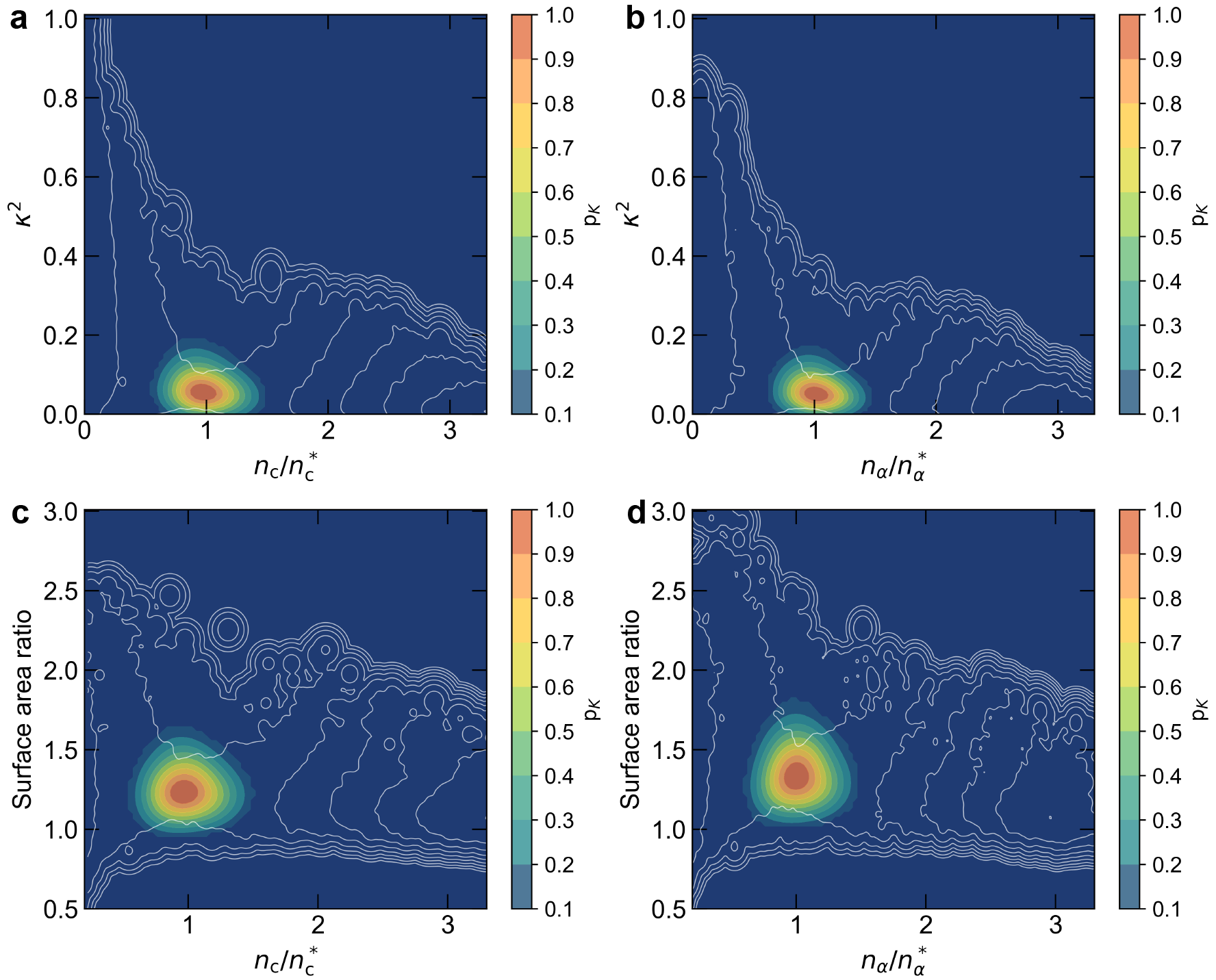}
        \caption{
        Kolmogorov distribution density in the plane of \textbf{a)} $n_\text{c}$, \textbf{b)} $n_\alpha$ and shape descriptor $\kappa^2$.
        Kolmogorov distribution density in the plane of \textbf{c)} $n_\text{c}$, \textbf{d)} $n_\alpha$ and surface area ratio ${A_{\text{cluster}}}/{A_{\text{sphere}}}$.
        The $x$ coordinates are rescaled according to the values of their critical numbers, respectively.
        The white isolines represent the reference free energy levels.}
        \label{sup_fig:shape_parameters}
    \end{figure}

\begin{figure}[h!]
        \centering
        \includegraphics[width=0.75
        \linewidth]{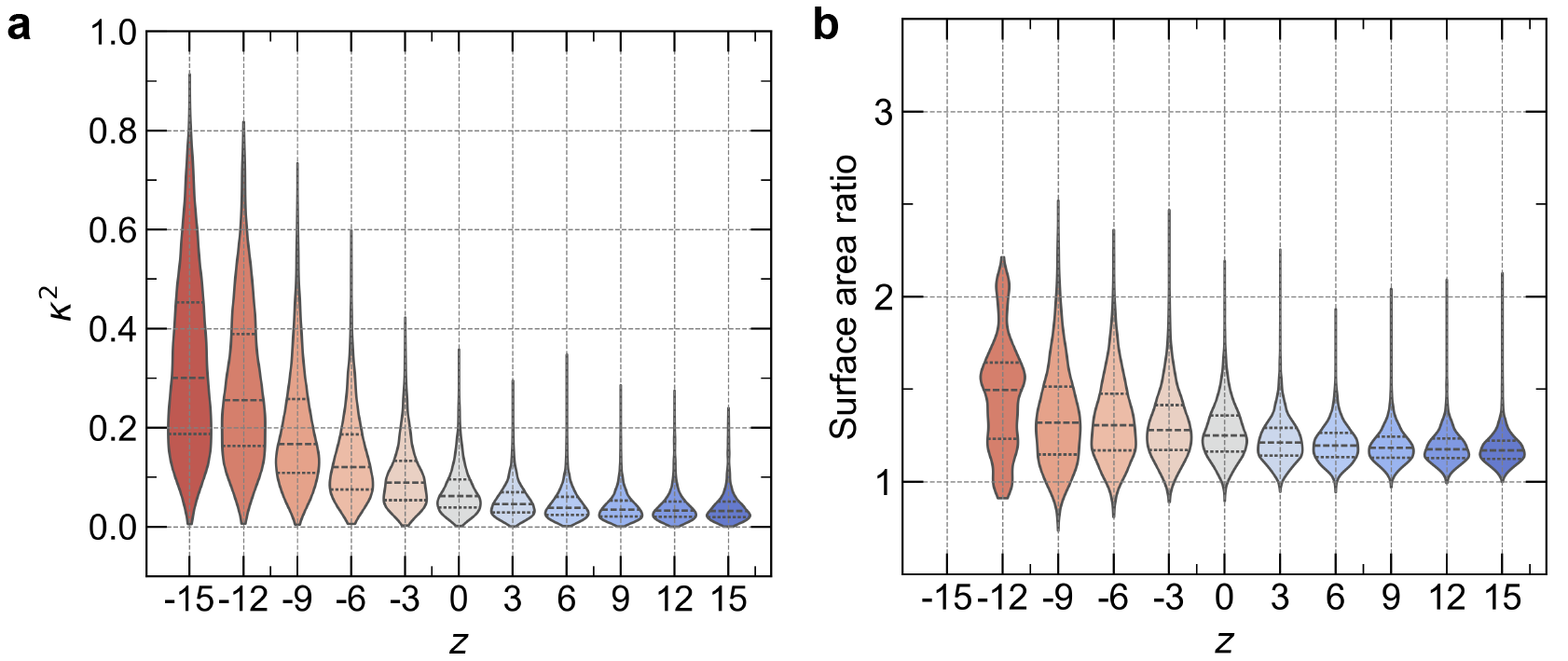}
        \caption{
        Violin plot of the distribution of \textbf{a)} shape descriptor $\kappa^2$ and \textbf{b)} surface area ratio ${A_{\text{cluster}}}/{A_{\text{sphere}}}$ of each bin along $z$.}
        \label{sup_fig:shape_parameters_violin}
    \end{figure}

    \yuanpeng{As shown in Fig.~\ref{sup_fig:shape_parameters}, the $p_\mathcal{K}$-based TSE reveals broad distributions for both shape descriptors. 
    The shape anisotropy $\kappa^2$ spans 0--0.2 across all projections. 
    The surface area ratio ranges from 1 to 1.65 when using conventional $n_\text{c}$, and from 1 to 1.75 with the probability-based $n_\alpha$.}
    This observation strengthens the finding that critical nuclei are not spherical.
    Furthermore, the violin plot of \yuanpeng{the $\kappa^2$ and the surface area ratio} as functions of $z$, shown in Fig.~\ref{sup_fig:shape_parameters_violin}, reveals the initial distribution of the asymmetry ratio along the reaction coordinate $z$, exhibiting a broad spread.
    This indicates significant shape fluctuations at the early stages of nucleation.
    As the nucleation progresses, this distribution narrows and shifts towards smaller values, reflecting a gradual transition towards more spherical clusters.

\begin{figure}[h!]
        \centering
        \includegraphics[width=0.8
        \linewidth]{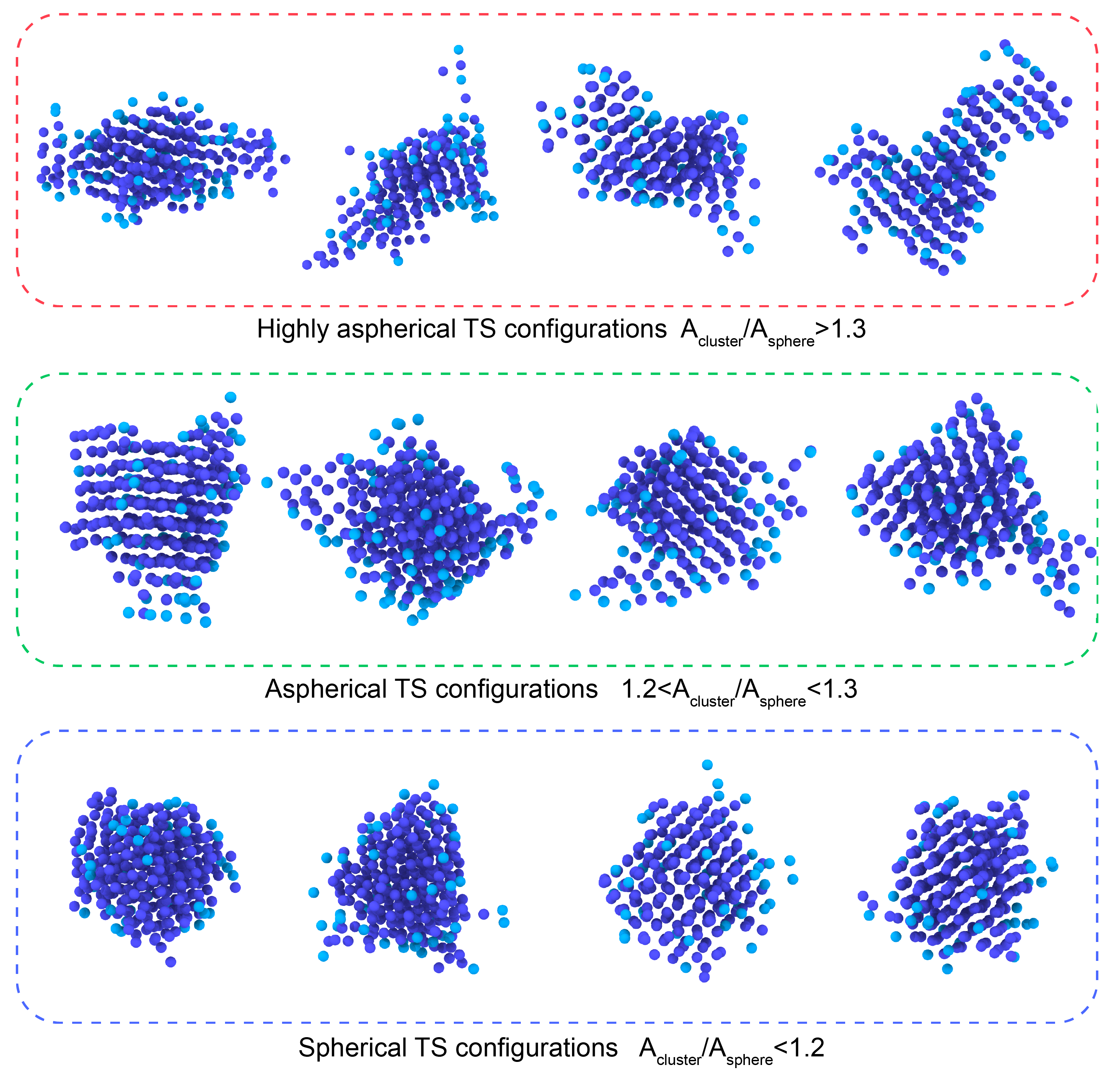}
        \caption{Snapshots of representative spherical, aspherical, and highly aspherical transition state (TS) configurations.
        Atoms forming the conventional $n_\text{c}$ nucleus are colored in blue, while the additional atoms included in the $n_\alpha$ definition are colored in cyan.}
        \label{sup_fig:snapshots_TS}
    \end{figure}

    To illustrate the diversity of nuclei shapes, we selected representative TS configurations and visualized them in Fig.~\ref{sup_fig:snapshots_TS}.
    \yuanpeng{These include spherical TS configurations (with $A_{\text{cluster}}/A_{\text{sphere}}< 1.2$ ), aspherical ones at the center of $p_\mathcal{K}$ (with $1.2 <A_{\text{cluster}}/A_{\text{sphere}} < 1.3$), and highly aspherical ones (with $A_{\text{cluster}}/A_{\text{sphere}}> 1.3$) by identifying clusters with $n_\text{c}$.}
    To highlight the distinction between $n_\text{c}$ and $n_\alpha$, atoms within the $n_\text{c}$ nucleus are colored blue, while additional atoms included in the $n_\alpha$ measure are colored cyan.

\clearpage
\newpage

\section{Probability-based $n_\alpha$}
    \label{sup_sec:probability_based_nalpha}

\begin{figure}[h!]
        \centering
        \includegraphics[width=0.55\linewidth]{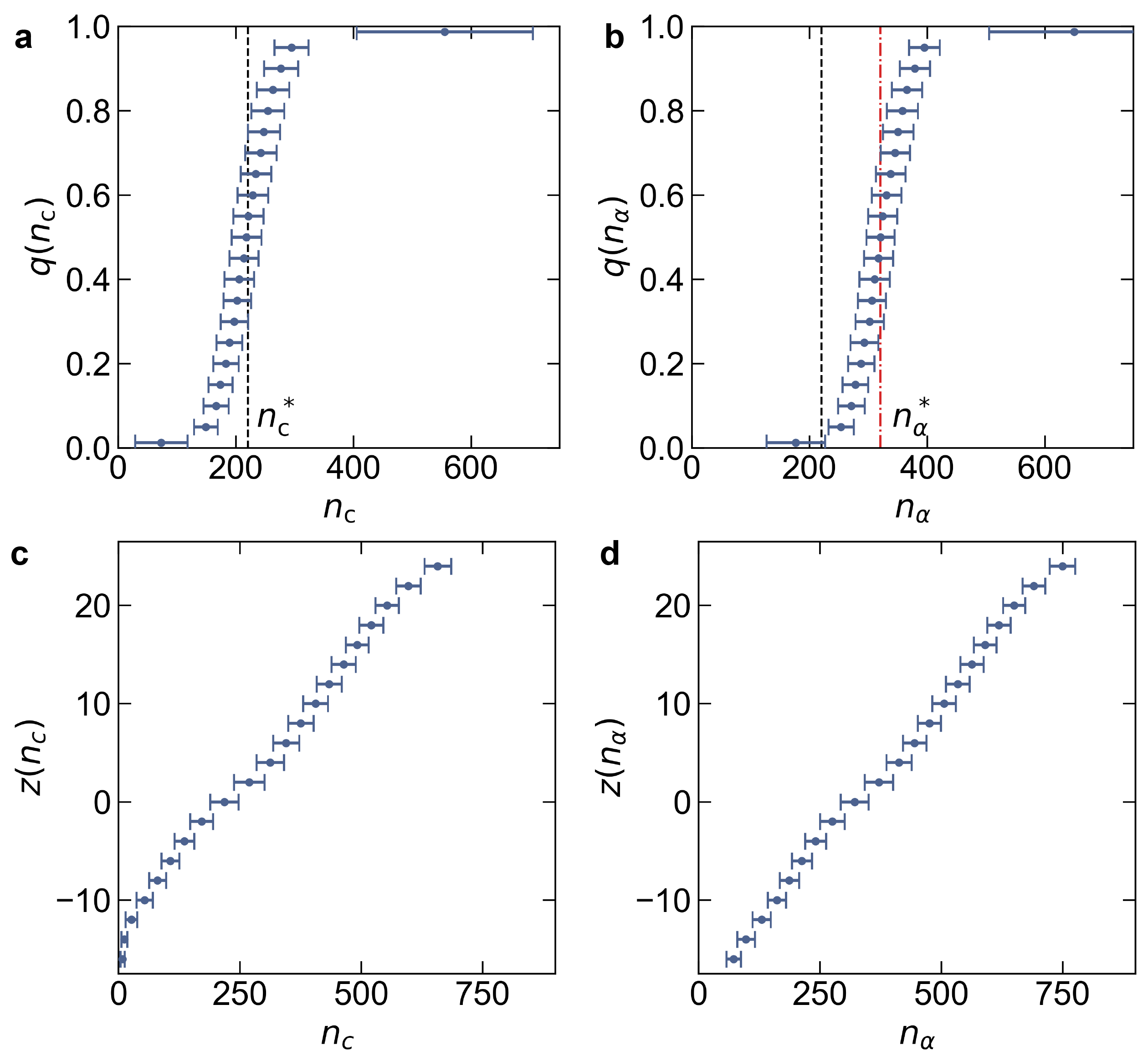}
        \caption{
        Projection of committor function $q$ as a function of \textbf{a)} $n_\text{c}$ and \textbf{b)} $n_\alpha$.
        Dashed lines referred to the critical nucleus size estimated from $q(n_\text{c}^*)=0.5$, while the red dashed-dotted line referred to the critical nucleus size estimated from $q(n_{\alpha}^*)=0.5$.
        Projection of $z$ as a function of \textbf{c)} $n_\text{c}$ and \textbf{d)} $n_\alpha$.
        }
        \label{sup_fig:qandz_nc_alpha}
    \end{figure}

\begin{figure}[h!]
        \centering
        \includegraphics[width=0.38\linewidth]{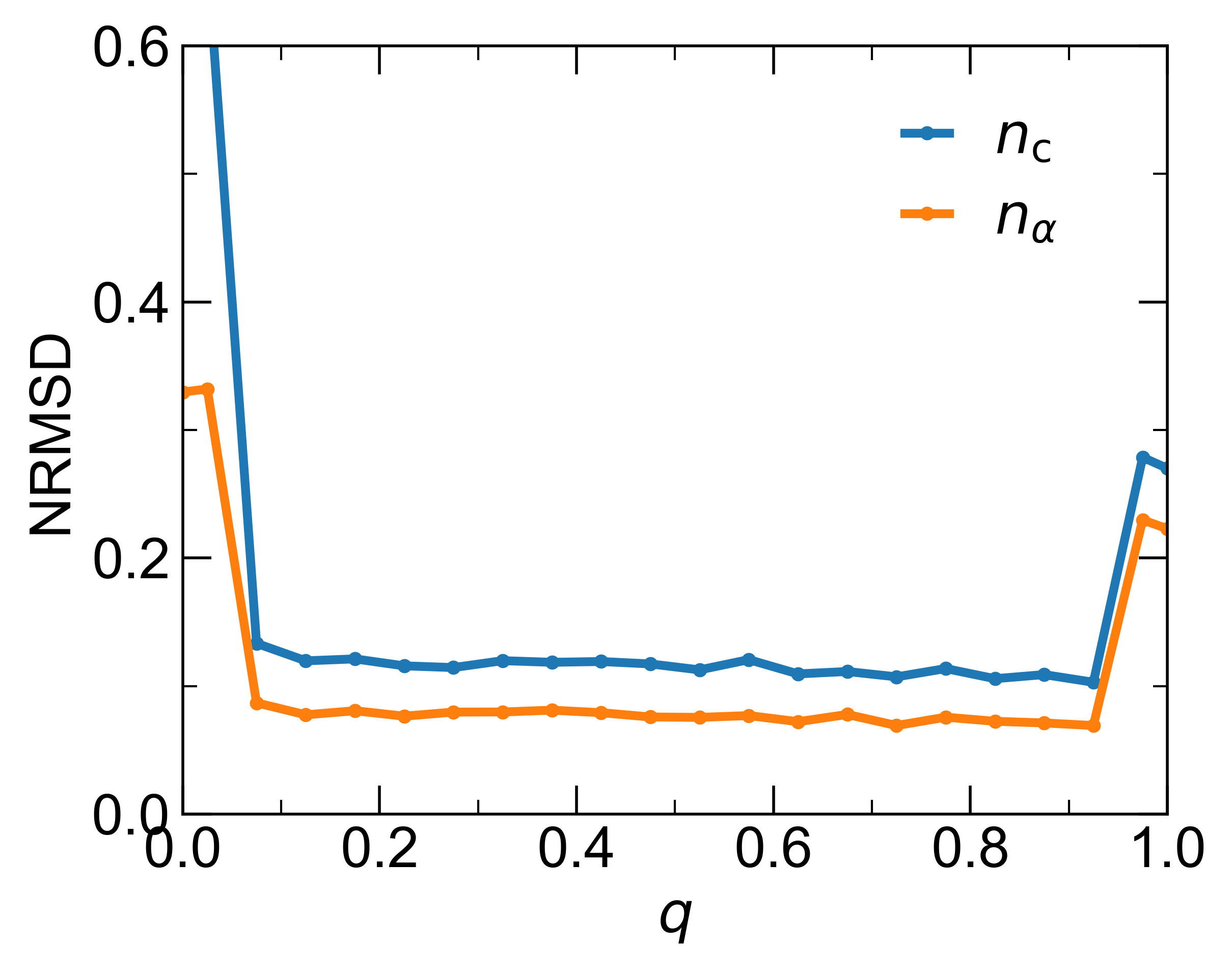}
        \caption{NRMSD of $n_\text{c}$ and $n_\alpha$ as functions of $q$.
        }
        \label{sup_fig:q_NRMSD}
    \end{figure}

    We further validate $n_\alpha$ as a descriptor of the nucleation transition by correlating it with the committor probability and comparing it against convetional $n_\text{c}$.
    We project the committor function $q(\mathbf{x})$ and $z(\mathbf{x})$ onto both $n_\text{c}$ and $n_\alpha$ metrics. 
    Fig.~\ref{sup_fig:qandz_nc_alpha} shows that $n_\alpha$ exhibits fluctuations comparable to those of $n_\text{c}$, suggesting that both descriptors yield similar fluctuation levels.  
    However, a balanced evaluation requires accounting for the number of atoms per bin under each criterion. 
    To quantify their differences, we compute the normalized root mean square deviation (NRMSD) along each bin of $q$ and $z$. 
    As shown in Fig.~\ref{fig:4}\textbf{b} and Fig.~\ref{sup_fig:q_NRMSD}, the $\alpha > \frac{1}{2}$ criterion minimizes NRMSD when normalized by the average atom count per bin, while capturing more atoms surrounding the emerging nucleus at comparable fluctuation levels. 
    This precision arises from $n_\alpha$\text{'}s probabilistic definition, which comprehensively includes all atoms in the solid-like peak, making it a more reliable and physically meaningful measure of nucleus size than $n_\text{c}$.
    To more clearly illustrate the effect of our probability-based $n_\alpha$, we highlight in Fig.~\ref{sup_fig:snapshots_TS} the atoms characterized by $n_\text{c}$ and those additionally included as part of the nucleus by $n_\alpha$ using distinct colors.

\clearpage
\newpage  

\section{Unbiased simulation of planer growing surfaces}
    \label{sup_sec:planer_growth}
    
\begin{figure}[h!]
        \centering
        \includegraphics[width=0.5\linewidth]{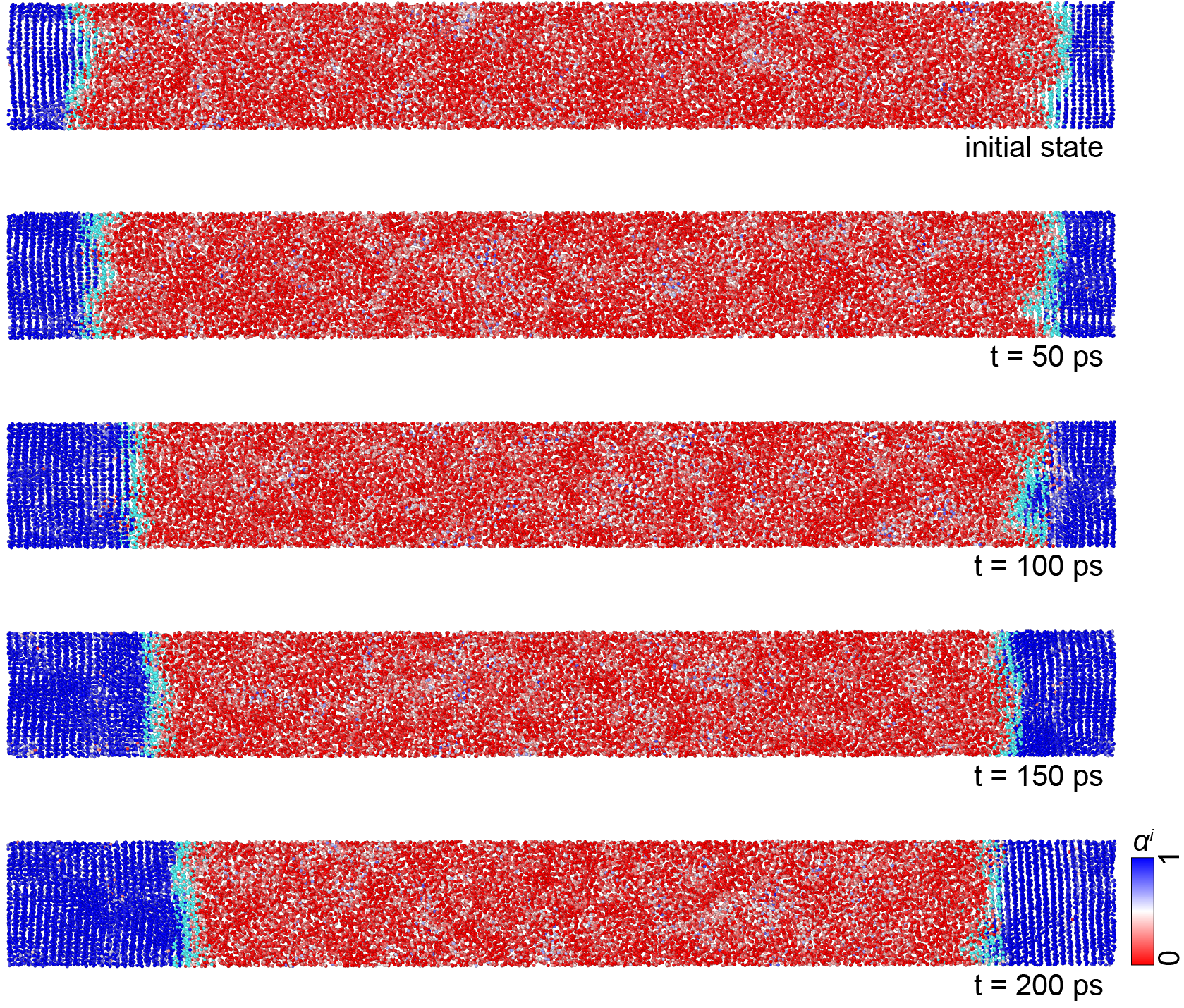}
        \caption{Snapshots illustrating the growth process at the planar interface between the liquid and the [111] solid lattice plane in an elongated liquid box, using unbiased molecular dynamics simulations.
        Atoms are colored according to their individual $\alpha^i$ values, with interfacial atoms highlighted in cyan for clarity.}
        \label{sup_fig:slabgrowth_new}
    \end{figure}

    \yuanpeng{To investigate the crystallization process in planer growing surfaces, we initialized unbiased MD simulations with crystallites positioned at the edges of a elongated liquid box. 
    Specifically, two crystallites with the size of $5 \times 5 \times 5$ nm$^3$ were placed at each end of the liquid box with the size of $50 \times 5 \times 5$ nm$^3$.
    The system was equilibrated for 20 ps before initiating the nucleation process.
    Simulations were conducted for different orientations of the solid-liquid interface, corresponding to the [100], [110], and [111] lattice planes.
    The results were averaged over five independent sets of nucleation processes, each run for 200 ps to observe the crystallization process along each lattice orientation.
    Snapshots illustrating the planer growth of crystallite is shown in Fig.~\ref{sup_fig:slabgrowth_new}.}
    
    \yuanpeng{To accurately locate the solid-liquid interface, we employed a systematic methodology based on the probability-based descriptor $\alpha^i$.
    First, we distinguished between crystal-like ($\alpha^i > \frac{1}{2}$) and liquid-like ($\alpha^i < \frac{1}{2}$) atoms.
    Crystal-like atoms that had at least one liquid-like atom within their first coordination shell were identified as interfacial atoms, forming the set $n_\text{int}$.
    The position of interface was determined by computing the spatial average $x$-coordinate, $\overline {x}$, of all $n_\text{int}$ atoms, where the $x$-axis aligns with the axial growth direction.
    To ensure the interface remain sufficiently planar, we evaluated the standard deviation of the $x^i$ coordinates of the $n_\text{int}$ atoms.
    Trajectories with this standard deviation exceeded $3.5 \, \text{\AA}$ were considered to have excessively non-planar interfaces and were excluded from further analysis.
    This filtering ensured a reliable characterization of the interface position.}

\begin{figure}[h!]
        \centering
        \includegraphics[width=0.5\linewidth]{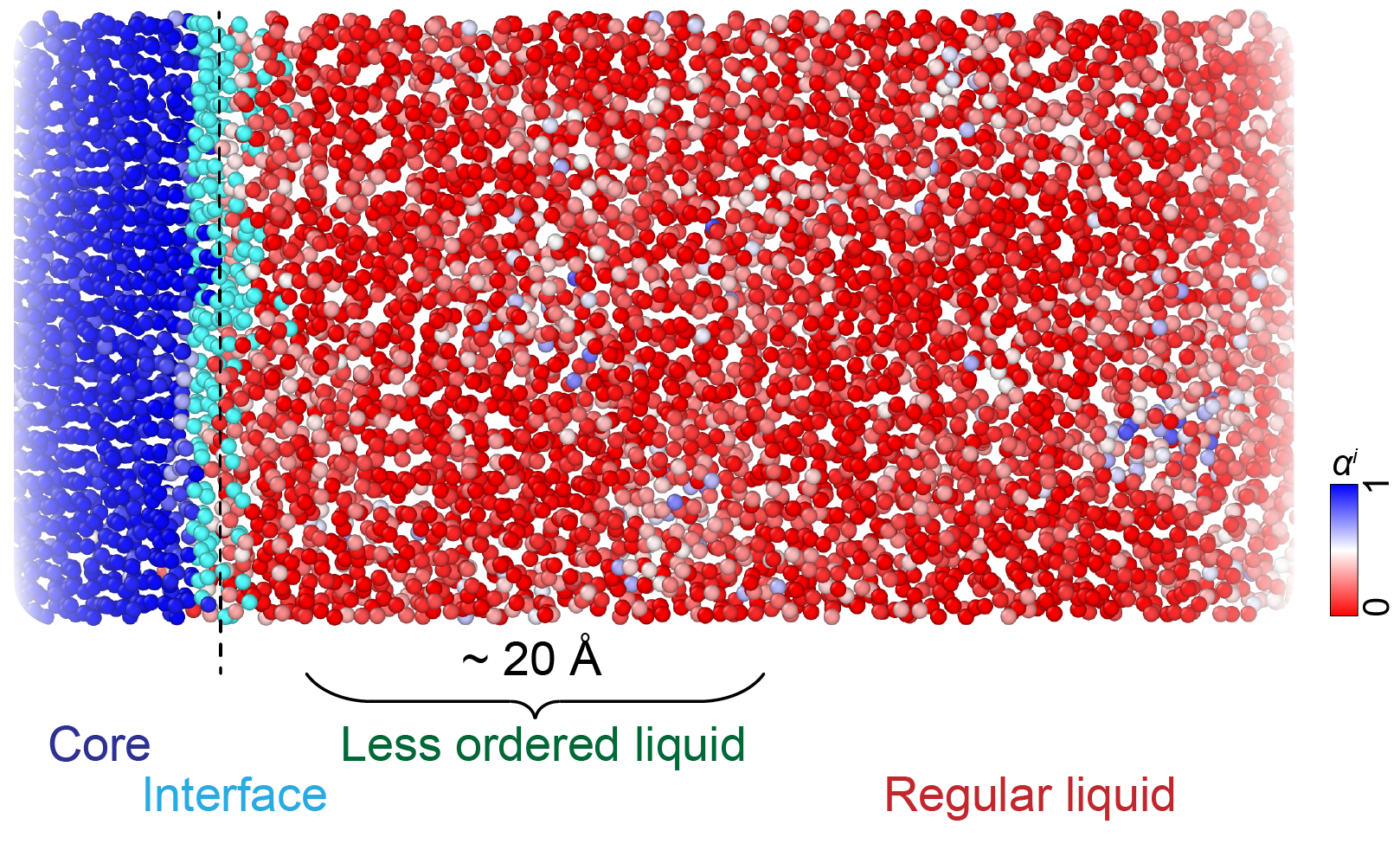}
        \caption{Schematic illustration of the atomic environment near the planar growing surface.
        A less-ordered liquid region with a thickness of $\sim20 \, \text{\AA}$ is present surrounding the solid-liquid interface.
        Atoms are colored based on their $\alpha^i$ values, with interfacial atoms highlighted in cyan for clarity.}
        \label{sup_fig:slabgrowth_schematic}
    \end{figure}

    \yuanpeng{To analyze the spatial variation of atomic properties near the solid-liquid interface, the simulation box was divided into bins along the $x$-axis, each with a width of $2.5 \, \text{\AA}$.
    The distance from the interface was defined as the distance from the center of each bin to the interface position.
    Atomic properties were computed for each bin and averaged over all trajectories that met the planarity criterion, as well as over the five independent simulations for each interface orientation.
    This approach enabled a detailed characterization of how atomic properties vary as functions of distance from the interface.
    A representative snapshot illustrating the atomic environment near the interface is provided in Fig.~\ref{sup_fig:slabgrowth_schematic}.}

\begin{figure}[h!]
        \centering
        \includegraphics[width=0.35\linewidth]{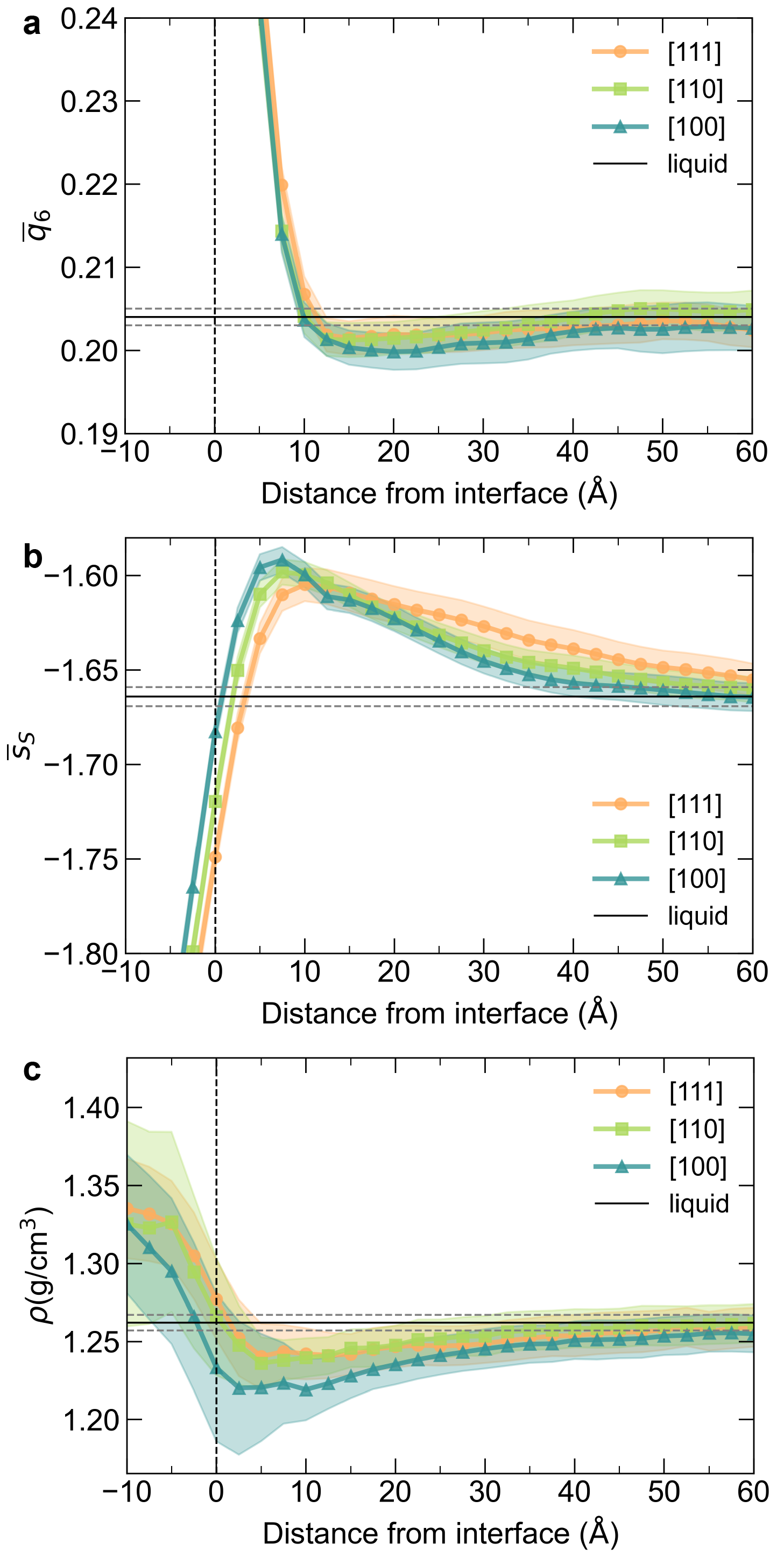}
        \caption{Variation of \textbf{a)} $\overline {q}_6$, \textbf{b)} $\overline s_{S}$, and \textbf{c)} $\rho$ as functions of distance from the interface.
        The dashed lines represent the standard deviation of these quantities in the bulk liquid.}
        \label{sup_fig:slab_additional}
    \end{figure}

    \yuanpeng{Similar to the trends observed in the reduction of the descriptor $\overline{\alpha}$ in Fig.~\ref{fig:5}, the bond order parameter $\overline{q}_6$, the locally averaged entropy $\overline{s}_S$, and the density $\rho$ all exhibit reduced values in a region near the solid-liquid interface compared to their values in the bulk liquid (see Fig.~\ref{sup_fig:slab_additional}).
    This observation further supports the presence of a less-ordered liquid region surrounding the solid-liquid interface, which plays a critical role in the nucleation process.}

\clearpage
\newpage

\section{Estimation of nucleation rate}
    \label{sup_sec:nucleation_rate}

    We first determine the self-diffusion coefficient $D$ of the undercooled liquid with the mean square displacement (MSD), defined as:

    \begin{equation}
        D=\frac{\text{MSD}}{6t}\text{ , where } \text{MSD}=\frac{1}{N}\sum_{i=1}^{N}[\vec{r_i}(t)-\vec{r_i}(0)]^2
        \label{eq:diffusion_coeff}
    \end{equation}

    \noindent where $\vec{r}_i(t)$ represents the position of particle $i$ at time $t$, and $N$ is the total number of particles.
    \yuanpeng{Next, to determine the effective friction coefficient $\gamma$ in overdamped Langevin dynamics corresponding to the diffusion coefficient obtained from molecular dynamics simulations, we use the Einstein relation:
    }
    
    \begin{equation}
        {\color{black} D=\frac{k_\text{B}T}{m\gamma}}
        \label{eq:einsten_relation}
    \end{equation}

    \noindent \yuanpeng{where $k_\text{B}$ is the Boltzmann constant, $T$ is the temperature, and $m$ is the mass per particle.
    By taking this relation, we obtain $\gamma \approx 3.3\,\text{ps}^{-1}$.}
    By fully exploring the metastable states and their reversible transitions, we achieve an optimal value for the functional $\mathcal{K}_m \approx 2.45\times 10^{-20}\, \textup{~\AA}^{-2}(\text{g/mol})^{-1}$, leading to a reaction rate of \yuanpeng{$\nu=4.67\times10^{-6}\,\text{s}^{-1}$} according to Eq.\ref{eq:reaction_rate_nu}.
    Furthermore, the transition rate from $A$ to $B$, $k_{A \to B}$, is expressed as,

    \begin{equation}
        k_{A \to B}=\frac{\nu}{\rho _A}
        \label{eq:kab}
    \end{equation}
    
    \noindent \yuanpeng{where $\rho _A=\lim_{T \to + \infty}\frac{T_A}{T}$ is the probability that the trajectory occupies state $A$, with $T_A$ being the total time spent in state $A$ and $T$ the total simulation time.
    This probability $\rho _A$ can be derived from the free energy difference of state $A$ and state $B$ as shown in Fig.~\ref{sup_fig:free_energy_difference},
    }
    
    \begin{equation}
        \frac{\rho _A}{\rho _B}=e^{-\beta \Delta G_{A-B}}
        \label{eq:rho_free_energy}
    \end{equation}

    \yuanpeng{
    Considering the large free energy difference of $\beta \Delta G_{A-B}\approx24.5$ in the undercooled system, we approximate $\rho_A \approx e^{-24.5} \approx 2.98 \times 10^{-11}$.
    Finally, the nucleation rate is calculated as,}
    
    \begin{equation}
        J = \frac{k_{A\to B}} {V}  
        \label{eq:nucleation_rate}
    \end{equation}

    \noindent \yuanpeng{where $V$ is the volume of the simulation box.
    We obtain $J  \approx 1.25 \times 10^{30} \, \text{m}^{-3} \text{s}^{-1}$, which is in much better agreement with the experimental result.}

    